\documentclass[aps,prb,twocolumn,floatfix]{revtex4}

\usepackage{epsfig,dcolumn}

\begin{document}
  \title{Thermodynamically consistent equilibrium
    properties of normal-liquid $^3$He}
  \author{M.\ Kollar$^*$}
  \author{D.\ Vollhardt}
  \affiliation{Theoretische Physik III, 
    Elektronische Korrelationen und Magnetismus,
    Institut f\"{u}r Physik,\\
    Universit\"{a}t Augsburg, 
    D-86135 Augsburg, Germany}
  \date{April 18, 2000; updated September 2, 2005}
  \begin{abstract}
    The high-precision data for the specific heat $C_{V}(T,V)$ of
    normal-liquid $^3$He obtained by Greywall, taken together with the
    molar volume $V(T_0,P)$ at one temperature $T_0$, are shown to
    contain the complete thermodynamic information about this phase in
    zero magnetic field.  This enables us to calculate the $T$ and
    $P$ dependence of all equilibrium properties of normal-liquid
    $^3$He in a thermodynamically consistent way for a wide range of
    parameters.  The results for the entropy $S(T,P)$, specific heat
    at constant pressure $C_P(T,P)$, molar volume $V(T,P)$,
    compressibility $\kappa(T,P)$, and thermal expansion coefficient
    $\alpha(T,P)$ are collected in the form of figures and tables.
    This provides the first complete set of thermodynamically
    consistent values of the equilibrium quantities of normal-liquid
    $^3$He. We find, for example, that $\alpha(T,P)$ has a
    surprisingly intricate pressure dependence at low temperatures,
    and that the curves $\alpha(T,P)$ vs $T$ do not cross at one
    single temperature for all pressures, in contrast to the curves
    presented in the comprehensive survey of helium by Wilks.
    
    \textbf{Corrected in cond-mat/9906222v3:} The sign of the
    coefficient $d_0$ was misprinted in Table~I of cond-mat/9906222v1
    and v2.  It now correctly reads $d_0=-7.1613436$. All results in
    the paper were obtained with the correct value of $d_0$. (We would
    like to thank for E.\ Collin, H.\ Godfrin, and Y.\ Bunkov for
    finding this misprint.)
  \end{abstract}
  \maketitle

  \section{Introduction}\label{intro}
  
  Liquid $^{3}$He is an exceptional system: at sufficiently low
  temperatures its normal phase is the prototype of a Landau Fermi
  liquid,\cite{Pines66,Baym78} while the superfluid phases, appearing
  at even lower temperatures, present examples of anisotropic BCS-type
  superfluids.\cite{Lee78,Vollhardt90} Therefore liquid $^{3}$He has
  been investigated systematically and in considerable detail at
  temperatures $T$ both below the critical temperature $T_c$ in the
  superfluid state $(T<T_c=0.0026\mbox{~K})$
  \cite{Wheatley75,Lee78} and in the Fermi-liquid regime $(T_c < T
  \alt 0.16 \mbox{~K})$.\cite{Wilks67,Wheatley70} At higher
  temperatures the dependence of the molar volume $V$, entropy $S$,
  thermal expansion coefficient $\alpha$, and specific heat $C_P$ on
  pressure $P$ had already been measured earlier by different
  groups.\cite{Brewer59a,Brewer59b,Sherman60,Lee61,Kerr62,Boghosian66}
  Their values are summarized in figures, as well as tables in the
  appendixes, in the comprehensive book by Wilks.\cite{Wilks67} Later
  measurements of some of these quantities were made, for example, by
  Grilly\cite{Grilly71} and by Abraham and Osborne.\cite{Abraham71}
  Subsequently, high-precision measurements of the specific heat at
  constant volume, $C_V(T,V)$, were performed in a wide temperature
  range by Greywall.\cite{Greywall83}
  
  In view of the fundamental importance of liquid $^3$He for our
  understanding of strongly correlated Fermi systems it would be
  desirable
   \begin{description}
   \item{(i)} to have complete knowledge of all equilibrium
     thermodynamic quantities in the full $(T,P)$ {\em and} $(T,V)$
     plane, and
   \item{(ii)} to be sure that these values indeed fulfill the usual
     thermodynamic relations, i.e., are {\em thermodynamically
       consistent}.
  \end{description}  
  In this paper we wish to point out that it is possible to {\em
    calculate} the full set of equilibrium thermodynamic quantities
  characterizing normal-liquid $^{3}$He from the specific heat
  $C_V(T,V)$ measured by Greywall,\cite{Greywall83} complemented by
  existing data for the molar volume $V(T_0,P)$ at a fixed temperature
  $T_0$ $=$ 0.1K as parametrized by
  Greywall.\cite{Greywall83,footnote1} Namely, these quantities are
  shown to contain the complete thermodynamic information about the
  system.  This enables us to calculate all equilibrium thermodynamic
  quantities of liquid $^3$He at zero magnetic field.  The advantage
  of this approach is that the temperature and pressure (or volume)
  dependences of these quantities are then guaranteed to be
  thermodynamically consistent. These results are collected into
  figures and tables and provide the first complete set of
  thermodynamically consistent values of the equilibrium quantities of
  normal-liquid $^3$He.
  
  In this paper we deal only with the normal-liquid phase of $^3$He.
  Nevertheless, since the measurements by Greywall\cite{Greywall83}
  were performed down to temperatures as low as 7~mK, we will present
  also limiting values for thermodynamic quantities in the limit
  $T\to0$, even though $^3$He would be superfluid then.  We note that
  Greywall later also published measurements of the specific heat and
  the melting curve of $^3$He based on a revised temperature
  scale.\cite{Greywall85,Greywall86} In order to maintain the
  thermodynamic consistency and since the new temperature scale leads
  to corrections\cite{Greywall85} in $C_V(T,V)$ of only $0.5\%$ above
  0.07~K and 1\% below 0.07~K, the calculations in this paper are
  based solely on the data of Ref.~\onlinecite{Greywall83}.
  
  The paper is organized as follows. In Sec.~\ref{info} it is shown
  how to obtain the complete thermodynamic information from the known
  experimental data. In Sec.~\ref{data} the calculation of
  thermodynamic quantities is explained in some detail. In
  Sec.~\ref{results} the molar volume $V(T,P)$, entropy $S(T,P)$,
  specific heat at constant pressure $C_P(T,P)$, compressibility
  $\kappa(T,P)$, and expansion coefficient $\alpha(T,P)$ are
  discussed, and are plotted and listed in tables.
  
  \section{Complete Thermodynamic Information}\label{info}
  
  A thermodynamic potential as a function of its natural variables
  contains the complete thermodynamic information about the system in
  equilibrium.  Here we show how to obtain the free energy $F(T,V)$ as
  a function of its natural variables $T$ and $V$, when the specific
  heat $C_V(T,V)$ and the molar volume $V(T_0,P)$ at a fixed
  temperature $T_0$ are known. Since $dF=-S\,dT-P\,dV$, and hence
  \begin{equation}
    F(T,V)=-\int\limits_{V_0}^V\!dV'\,P(T_0,V')-
    \int\limits_{T_0}^T\!dT'\,S(T',V)+\mbox{const},\label{ftv}
  \end{equation}
  one needs the entropy $S(T,V)$ and the pressure $P(T_0,V)$ to
  calculate $F(T,V)$.  The entropy is obtained from $C_V(T,V)$ by
  integration:
  \begin{equation}\label{stv}
    S(T,V)=\int\limits_0^T\!\frac{dT'}{T'}\,C_V(T',V),
  \end{equation}
  and $P(T_0,V)$ can, in principle, be calculated from the molar
  volume $V(T_0,P)$ by inversion.  In the following we will not
  calculate the free energy explicitly according to Eq.~(\ref{ftv}).
  Instead, we make use of the Maxwell relation
  \begin{equation}\label{ps-maxrel}
    \left(\frac{\partial{P}}{\partial{T}}\right)_{\!{V}}
    =\left(\frac{\partial{S}}{\partial{V}}\right)_{\!{T}}
  \end{equation}
  to obtain the pressure
  \begin{equation}\label{ptv}
    P(T,V)=P(T_0,V)+\int\limits_{T_0}^T\!dT'\,
    \left(\frac{\partial{S(T',V)}}{\partial{V}}\right)_{\!{T'}}
    ,
  \end{equation}
  and hence by inversion the molar volume $V(T,P)$.
  Equation~(\ref{ptv}) contains a central observation of this paper:
  $P(T_0,V)$ or $V(T_0,P)$ at a single temperature can be extended to
  all temperatures if $C_V(T,V)$ is known.
  
  From these equations thermodynamic quantities such as the entropy
  $S(T,P)$ as well as derivatives of $V(T,P)$, such as the
  compressibility $\kappa$ and the expansion coefficient $\alpha$, can
  be calculated as functions of pressure (instead of volume) and
  temperature.
  
  \section{Calculation of Thermodynamic Quantities}\label{data}
  
  Based on his measurements, Greywall\cite{Greywall83} provided
  interpolation formulas $c_1(T,V)$, $c_2(T,V)$, and $v_0(P)$,
  representing the specific heat at constant volume $C_V(T,V)$
  \begin{equation}
    \frac{1}{R}\,C_V(T,V)=\left\{
      \begin{array}{ll}
        c_1(T,V)
        &\mbox{ for }T<T_0\\
        c_2(T,V)
        &\mbox{ for }T\geq T_0
      \end{array}
    \right.\label{cgleichung}
    , 
  \end{equation}
  as well as the molar volume $V(T_0,P)$ at temperature $T_0=0.1$~K
  \begin{equation}
    V(T_0,P)=v_0(P)\label{v0gleichung}
    .
  \end{equation}
  Greywall fitted the interpolation formulas in the temperature range
  $0.007 \leq T \leq 2.5\mbox{~K}$ for molar volumes $26.169 \leq V
  \leq 36.820\mbox{ cm}^3$.  His interpolation formulas are given by
  \begin{eqnarray}
    c_1(T,V)&=&
    \sum_{i=1}^{5}
    \sum_{j=0}^{3}
    a_{ij}
    \frac{T^i}{V^j},
    \label{c1def}
    \\
    c_2(T,V)&=&
    \sum_{i=0}^{3}
    \sum_{j=0}^{2}
    \left[
      b_{ij}+c_{ij}\,e^{-d(V)/T}
    \right]
    \frac{V^j}{T^i},
    \label{c2def}
    \\
    v_0(P)&=&
    \sum_{i=0}^{5}
    a_i\,P^i,
    \label{v0def}
  \end{eqnarray}
  with $d(V)=\sum_{j=0}^{2}d_j\,V^j$.  Here and in the following $T$,
  $V$ and $P$ are in units of K, cm$^3$, and bar, respectively.
  Below we will also need the inverse of Eq.~(\ref{v0gleichung}),
  i.e., $p_0(V)=P(T_0,V)$. We use the function
  \begin{equation}
    \label{p0def}
    p_0(V)=
    \sum_{i=1}^{7}
    b_i\,(V-a_0)^i,
  \end{equation}
  which we fitted to the inverse of $v_0(P)$ by a least-square fit
  (rms deviation 0.2 bar).\cite{footnote2} The parameters $b_i$, as
  well as Greywall's parameters appearing in
  Eqs.~(\ref{c1def})-(\ref{v0def}), are listed in Table
  \ref{paramtable}.
    
  Using Eqs.~(\ref{stv}) and (\ref{ptv}) together with
  Eq.~(\ref{cgleichung}) we obtain the entropy $S(T,V)$ and pressure
  $P(T,V)$ as
  \begin{widetext}
  \begin{eqnarray}
    \frac{1}{R}\,S(T,V)&=&
    \left\{
      \begin{array}{ll}
        s_1(T,V)
        &\mbox{ for }T<T_0
        \\
        s_2(T,V)+s_1(T_0,V)-s_2(T_0,V)
        &\mbox{ for }T\geq T_0
      \end{array}
    \right.,
    \label{sgleichung}
    \\
    P(T,V)&=&\left\{
      \begin{array}{ll}
        p_0(V)+q\,[p_1(T,V)-p_1(T_0,V)]
        &\mbox{ for } T<T_0\\
        p_0(V)+q\,[p_2(T,V)-p_2(T_0,V)]\\
        \;\;+(T-T_0)\,q\,
        \frac{\partial }{\partial
          V}\,{[s_1(T_0,V)-s_2(T_0,V)]}
        &\mbox{ for }T\geq T_0
      \end{array}
    \label{pgleichung}
    \right.,
  \end{eqnarray}
  where the auxiliary functions $s_i(T,V)$ and $p_i(T,V)$ are defined
  for $i=1,2$ via ${\partial s_i}/{\partial T}=c_i/T$, ${\partial
    s_i}/{\partial V}={\partial p_i}/{\partial T}$, with $s_1(0,V)=0$,
  and $q$ is a conversion factor, $q=83.1451$.  After analytically
  performing the necessary integrations we obtain the following
  expressions:
  \begin{eqnarray}
    s_1(T,V)&=&
    \sum_{i=1}^{5}
    \sum_{j=0}^{3}
    \frac{a_{ij}}{i}
    \frac{T^i}{V^j},
    \\
    p_1(T,V)&=&
    -
    \sum_{i=1}^{5}
    \sum_{j=0}^{3}
    \frac{j\,a_{ij}}{i(i+1)}
    \frac{T^{i+1}}{V^{j+1}},
    \\
    s_2(T,V)&=&
    \sum_{j=0}^{2}
    \left[
      b_{0j}\ln T-
      \sum_{i=1}^{3} 
      \frac{b_{ij}}{i}
      \frac{1}{T^i}
      +
      \sum_{i=0}^{2} 
      \sum_{k=i+1}^{3} 
      \frac{(k-1)!\,c_{kj}}{i!\,d(V)^{k-i}\,T^i}
      e^{-d(V)/T}
    \right]V^j,
    \\
    p_2(T,V)&=&
    \sum_{j=1}^{2}
    \left\{
      (b_{0j}T-b_{1j})\ln T-b_{0j}T+
      \sum_{i=2}^{3}\frac{b_{ij}}{i(i-1)}\frac{1}{T^{i-1}}
    \right.
    \nonumber\\
    &&
    +\left.
    \sum_{k=1}^{3}
    \frac{(k-1)!\,c_{kj}}{d(V)^{k-1}}
    \left[
      \left(
        \frac{T}{d(V)}+\frac{1}{2}\delta_{k3}
      \right)
      e^{-d(V)/T}     
      +
      \delta_{k1}\,
      \mbox{Ei}\!\left(-\frac{d(V)}{T}\right)
    \right]
    \right\}
    j\,V^{j-1}
    \nonumber\\
    &&+
    \sum_{j=0}^{2}
    \sum_{k=1}^{3}
    \frac{(k-1)!\,c_{kj}}{d(V)^k}
    \left[
      1-k\left(1+\frac{T}{d(V)}\right)-\delta_{k3}\frac{d(V)}{2T}
    \right]
      e^{-d(V)/T} 
      d'(V)V^{j}.
  \end{eqnarray}
  \end{widetext}
  The only nonelementary function that appears in these expressions
  is the exponential integral $\mbox{Ei}(x)$, defined
  by
  \begin{equation}
    \label{eidef}
    \mbox{Ei}(-x)=-
    \int\limits_{x}^{\infty}\frac{dt}{t}\,e^{-t},\;\;\;\;x>0.
  \end{equation}
  This function can be calculated numerically without
  difficulty.\cite{Press92}
  
  Other thermodynamic quantities can be calculated from
  Eqs.~(\ref{cgleichung}), (\ref{sgleichung}), and (\ref{pgleichung}).
  The specific heat at constant pressure $C_P$ is given by
  \begin{equation}\label{cpcv}
    C_P(T,P)=T
    \left(\frac{\partial{S}}{\partial{T}}\right)_{\!{P}}
    =T
    \left(\frac{\partial{S}}{\partial{T}}\right)_{\!{V}}
    -T\frac{
      \left(\frac{\partial{P}}{\partial{T}}\right)_{\!{V}}^2
      }{
      \left(\frac{\partial{P}}{\partial{V}}\right)_{\!{T}}      
      },
  \end{equation}
  and the isothermal compressibility and isobaric expansion
  coefficient are calculated from Eq.~(\ref{pgleichung}) as
  \begin{eqnarray}
    \label{kappadef}
    \kappa(T,P)&=&-\frac{1}{V}
    \left(\frac{\partial{V}}{\partial{P}}\right)_{\!{T}}    
    =-\frac{1}{V}
    \left(\frac{\partial{P}}{\partial{V}}\right)_{\!{T}}^{-1},
    \\
    \label{alphadef}
    \alpha(T,P)&=&\frac{1}{V}
    \left(\frac{\partial{V}}{\partial{T}}\right)_{\!{P}}    
    =\kappa
    \left(\frac{\partial{P}}{\partial{T}}\right)_{\!{V}}
    .
  \end{eqnarray}
  After obtaining $V(T,P)$ from Eq.~(\ref{pgleichung}) by inversion,
  these quantities can be calculated as a function of $T$ and $P$.  It
  is also possible to calculate higher derivatives in terms of known
  functions by repeatedly applying standard thermodynamic relations,
  e.g., 
  \begin{widetext}
  \begin{eqnarray}
    \left(\frac{\partial{\kappa}}{\partial{P}}\right)_{\!{T}}
    &=&
    \kappa^2-V^2\kappa^3
    \left(\frac{\partial^2{P}}{\partial{V}^2}\right)_{\!{T}}    
    ,
    \label{firstderivs}
    \\
    \left(\frac{\partial{\kappa}}{\partial{T}}\right)_{\!{P}}
    &=&
    -\kappa\,\alpha
    +V\kappa^2\alpha
    \left(\frac{\partial^2{P}}{\partial{V}^2}\right)_{\!{T}}    
    +V\kappa^2
    \left(\frac{\partial^2{S}}{\partial{V}^2}\right)_{\!{T}}    
    ,
    \\
    \left(\frac{\partial{\alpha}}{\partial{P}}\right)_{\!{T}}
    &=&
    -\left(\frac{\partial{\kappa}}{\partial{T}}\right)_{\!{P}}    
    \label{dkdt-dadp},
    \\
    \left(\frac{\partial{\alpha}}{\partial{T}}\right)_{\!{P}}
    &=&
    -\alpha^2
    +V^2\kappa\,\alpha^2
    \left(\frac{\partial^2{P}}{\partial{V}^2}\right)_{\!{T}}    
    +2\,V\kappa\,\alpha
    \left(\frac{\partial^2{S}}{\partial{V}^2}\right)_{\!{T}}    
    +\kappa
    \left(\frac{\partial^2{P}}{\partial{T}^2}\right)_{\!{V}}
    .
    \label{lastderivs}
  \end{eqnarray}  
  \end{widetext}
  Of these higher derivatives we will only evaluate the pressure
  dependence of
  $\left(\frac{\partial{\alpha}}{\partial{T}}\right)_{\!{P}}$ in the
  limit $T\to0$.

  \section{Results and Discussion}\label{results}
  
  In this section we will present both figures and tables of
  thermodynamic quantities calculated by the procedure described
  above. The allowed temperature range for these interpolation
  formulas is $0\leq T\leq2.5\mbox{~K}$, where $T=0$ denotes the
  extrapolation of the Fermi liquid data at $T=7~\mbox{mK}$ to zero
  temperature, disregarding the occurrence of superfluid phases.  The
  allowed range of molar volumes is $26.16 \leq V \leq 36.85\mbox{
    cm}^3$.  Note that for low molar volumes the pressure is always
  below the minimum of the melting
  pressure\cite{Greywall82,Greywall83} at $P\approx29.3\mbox{~bar}$
  and $T\approx0.32\mbox{~K}$. For high molar volumes the volume range
  includes the point $V(T=0,P=0)\approx36.85\mbox{~cm}^3$, this
  representing a slight extrapolation of Greywall's original formulas
  which extended only to $V=36.82\mbox{~cm}^3$. The resulting range of
  pressures is $0\leq P\alt28\mbox{-}29\mbox{~bar}$, depending on
  temperature.
  
  Below we present results for the molar volume, entropy, and specific
  heat as a function of pressure and temperature, as well as the first
  derivatives of the molar volume, the compressibility, and the
  thermal expansion coefficient.\cite{footnote3} Using
  Eqs.~(\ref{firstderivs})-(\ref{lastderivs}) of Sec.~\ref{data} it is
  also possible, in principle, to calculate higher derivatives, but
  because their reliability is difficult to judge we will calculate
  only the slope of the expansion coefficient as a function of
  pressure for $T\to0$.
  
  For pressures 0, 5, 10, 15, 20, 25, and 28 bar Tables
  \ref{firsttable}-\ref{lasttable} show data vs $T$ at temperatures
  $0\leq T \leq 2.5$~K. Table \ref{zerotemptable} lists data vs $P$
  at $T=0$.  We note that, for $P=0$, calculations are restricted to
  $T<1\mbox{~K}$ since at higher temperatures the calculated molar
  volumes $V(T,P=0)$ become larger than $36.85$ cm$^3$ and are thus
  outside the regime of the interpolation formulas.  We also note that
  since the interpolation formulas are given by two different
  expressions for temperatures above and below $T=0.1\mbox{~K}$, there
  appear weak cusps or discontinuities in some of the curves at this
  temperature, which we purposely did not smooth out.
    
  \subsection{Molar volume $V(T,P)$}
  
\begin{figure}[p]
    \centerline{\epsfig{width=\columnwidth,file=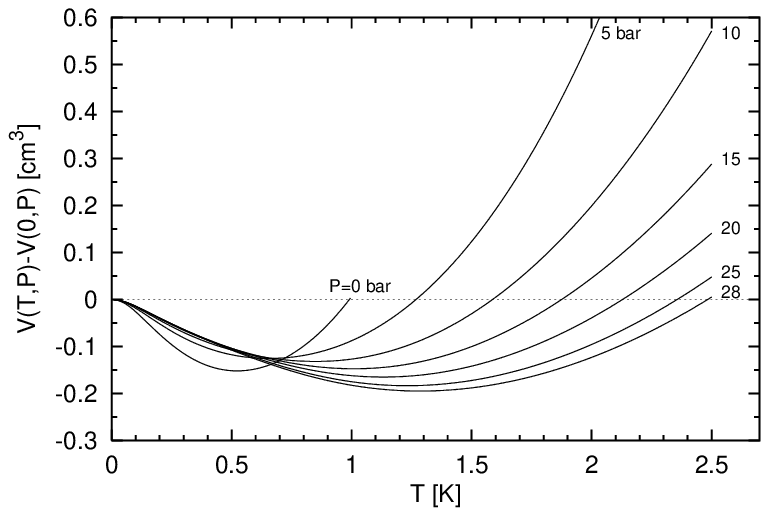}}
    \caption{Molar volume $V(T,P)$ vs $T$ at several pressures $P$,
      plotted relative to its value at $T=0$ shown in
      Fig.~\ref{v-vs-p}.\label{v-vs-t-p}}
  \end{figure}

\begin{figure}[p]
    \centerline{\epsfig{width=\columnwidth,file=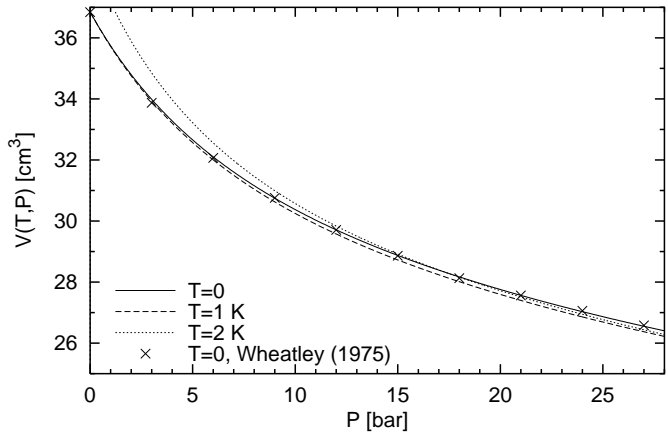}}
    \caption{Molar volume $V(T,P)$ vs $P$ at temperatures
      $T=0,$ 1, 2 K.  The data of Wheatley
      (ref.~\protect\onlinecite{Wheatley75}) at $T=0$ are also
      shown.\label{v-vs-p}}
  \end{figure}

  The temperature dependence of the molar volume $V(T,P)$ is plotted
  in Fig.~\ref{v-vs-t-p} relative to its zero-temperature value.  The
  pressure dependence of $V(T,P)$ is shown in Fig.~\ref{v-vs-p} for
  $T=0,1,2~\mbox{K}$. The change in volume over the temperature range
  $0\mbox{-}2.5$ K is small ($\alt4\%$).
  
  At low temperatures the volume is seen to {\em decrease} upon
  increase of temperature, implying a negative thermal expansion
  coefficient. This is found in many strongly correlated fermion
  systems and is due to the anomalous {\em increase} of entropy with
  pressure in these systems as will be discussed in the following two
  subsections.
  
  In the limit $T\to0$ we compared our results for $V(0,P)$ (see also
  Table \ref{zerotemptable}) with the values determined by
  Wheatley.\cite{Wheatley75} We find very good agreement for all
  pressures, the relative difference between these values being less
  than 0.5\%.  At higher temperatures ($1 \leq T\leq2.4$~K)
  our evaluation of $V(T,P)$ can be compared to the corrected data of
  Sherman and Edeskuty,\cite{Sherman60} which are tabulated in Wilks'
  book.\cite{Wilks67} Again the agreement is very good for all
  pressures, with relative differences less than 0.6\%.  Comparison
  with the data of Abraham and Osborne\cite{Abraham71} also shows very
  good agreement, with relative differences of at most 0.3\%.
  
  \subsection{Entropy $S(T,P)$ and specific heat $C_P(T,P)$}
   
\begin{figure}[p]
    \centerline{\epsfig{width=\columnwidth,file=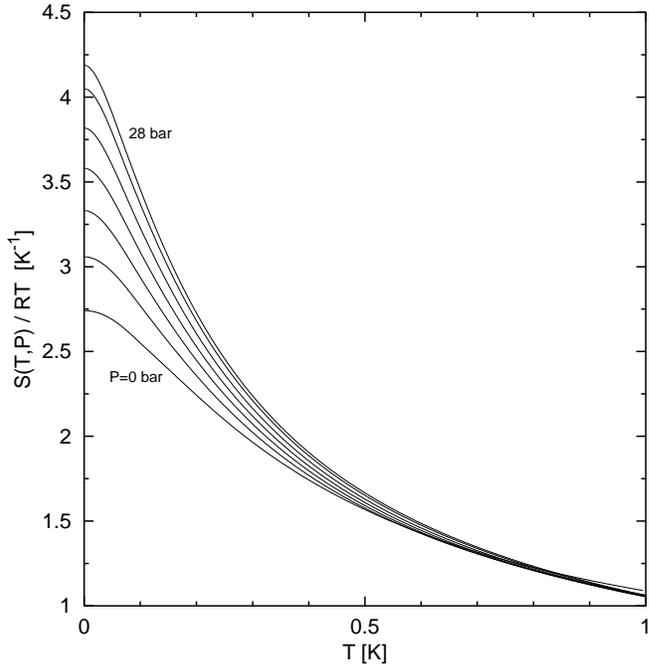}}
    \caption{Entropy $S(T,P)$ divided by temperature vs $T$ 
      at pressures $P=0,$ 5, 10, 15, 20, 25, 28 bar.\label{s-vs-t}}
  \end{figure}

\begin{figure}[p]
    \centerline{\epsfig{width=\columnwidth,file=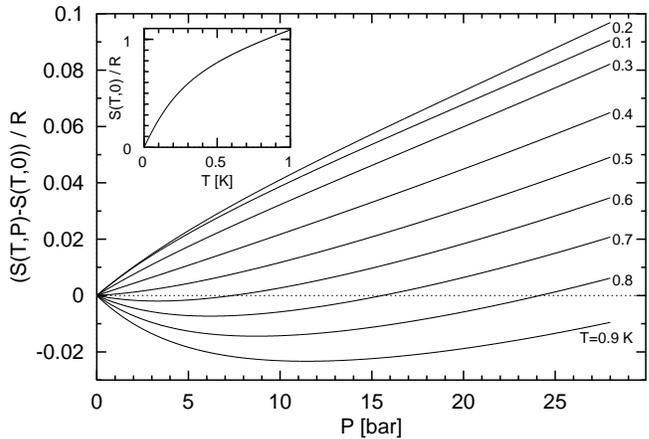}}
    \caption{Entropy $S(T,P)$ vs $P$ at several
      temperatures $T$, plotted relative to its value at $P=0$
      (inset).\label{s-vs-p-rel}}
  \end{figure}

  In Fig.~\ref{s-vs-t} $S(T,P)/T$ is shown as a function of
  temperature for several values of pressure. In Fig.~\ref{s-vs-p-rel}
  $S(T,P)$ is plotted as a function of pressure, relative to its value
  at $P=0$ shown as an inset.  Comparison of
  Fig.~\ref{s-vs-p-rel} with earlier, independent
  measurements\cite{Brewer59b,Lee61} of the entropy shows good
  quantitative agreement.
  
\begin{figure}[p]
    \centerline{\epsfig{width=\columnwidth,file=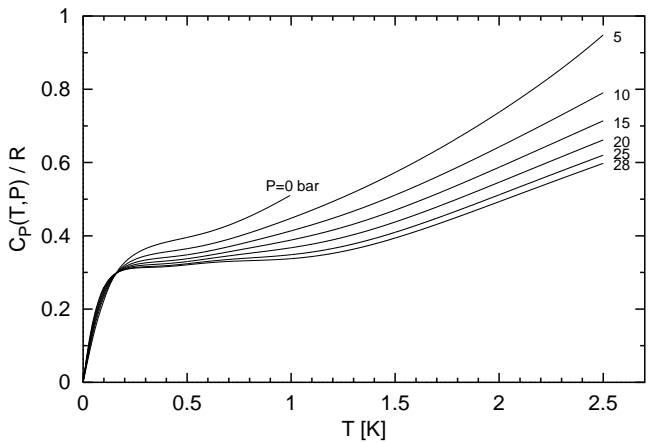}}
    \caption{High-temperature behavior of the specific heat at
      constant pressure $C_P(T,P)$.\label{c-vs-t}}
  \end{figure}

  The specific heat $C_P(T,P)$ is plotted as a function of temperature
  in Fig.~\ref{ct-vs-t} and \ref{c-vs-t}, and as a function of
  pressure in Fig.~\ref{ct-vs-p}.  As expected, the difference between
  the values of the specific heat at constant pressure $C_P$ and the
  previously published\cite{Greywall83} data at constant volume $C_V$
  is very small for low temperatures.  However, at the highest
  available temperature, $T=2.5\mbox{~K}$, this difference increases
  to about 10\%.\cite{footnote4} 
  
\begin{figure}[p]
    \centerline{\epsfig{width=\columnwidth,file=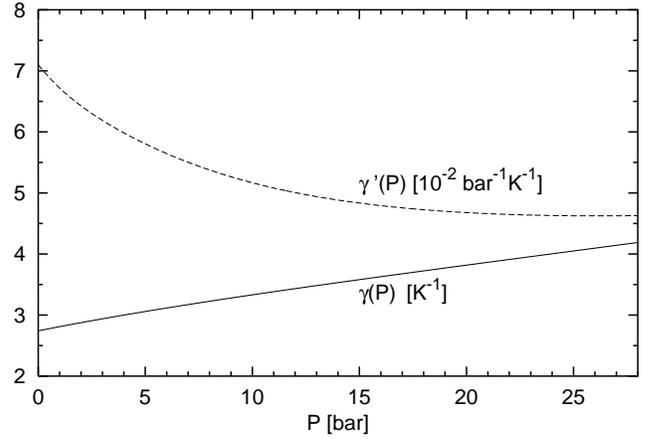}}
    \caption{Linear coefficient $\gamma(P)$ of the specific heat 
      [Eq.~(\ref{gammadef})] vs $P$, and its derivative with respect
      to $P$, $\gamma'(P)$.\label{g-vs-p}}
  \end{figure}

\begin{figure}[p]
    \centerline{\epsfig{width=\columnwidth,file=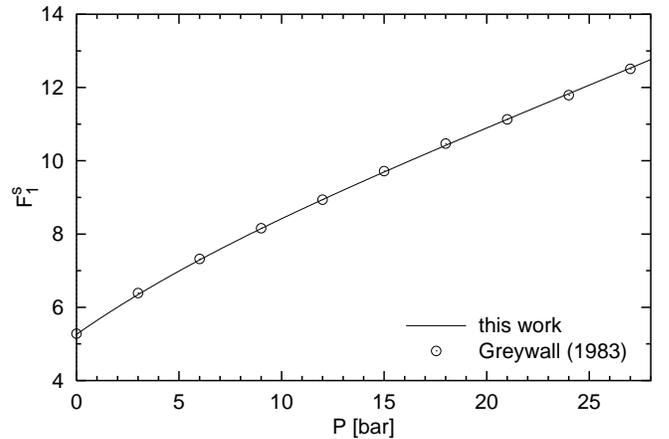}}
    \caption{Fermi liquid parameter $F_1^s$ vs $P$, 
      compared to the data obtained by Greywall
      (Ref.~\protect\onlinecite{Greywall83})\label{f1-vs-p}}
  \end{figure}

  At very low temperatures $^3$He is a Landau Fermi
  liquid\cite{Baym78} with an entropy and specific heat linear in $T$,
  \begin{equation}\label{gammadef}
    S(T,P)=C_P(T,P)=\gamma(P)RT,\;\;\;T\to0.
  \end{equation}
  The coefficient $\gamma$ can be expressed in terms of the effective
  mass $m^*$ or the Landau
  parameter $F_1^s$ as
  \begin{equation}\label{gammalp}
    \frac{\gamma}{\gamma_0}=\frac{m^*}{m}=1+\frac{1}{3}F_1^s,
  \end{equation}
  where\cite{Wheatley75}
  \begin{eqnarray}
    \label{gamma0}
    \gamma_0
    &=&
    \frac{\pi^2k_Bm}{\hbar^2}  
    \left(\frac{V(0,P)}{3\pi^2N}\right)^{\frac{2}{3}} 
    \nonumber\\
    &=&
    8.991\cdot10^{-2}\mbox{K}^{-1}
    \left(\frac{V(0,P)}{\mbox{cm}^3}\right)^{\frac{2}{3}}
  \end{eqnarray}
  is the corresponding coefficient of a hypothetical free fermion gas
  with atomic mass of $^3$He, $m=5.009\cdot10^{-24}\mbox{g}$, at the
  same density. The behavior of $\gamma(P)$ and its derivative with
  respect to $P$ are shown in Fig.~\ref{g-vs-p}.  The Fermi liquid
  parameter $F_1^s$ is plotted in Fig.~\ref{f1-vs-p} and compared to
  the values obtained by Greywall\cite{Greywall83} from $\gamma(V)$
  using $V(0.1\mbox{~K},P)$ instead of $V(0,P)$ in Eq.~(\ref{gamma0}).
  The agreement is excellent, because the difference between
  $V(0.1\mbox{~K},P)$ and $V(0,P)$ is small.
  
  The coefficient $\gamma(P)$ is seen to increase with pressure (see
  Fig.~\ref{g-vs-p}), implying that, at low temperatures, both the
  entropy $S(T,P)$ and the specific heat $C_P(T,P)$ increase with
  pressure: $\left(\frac{\partial{S}}{\partial{P}}\right)_{\!{T}}>0$,
  $\left(\frac{\partial{C_P}}{\partial{P}}\right)_{\!{T}}>0$. This may
  be attributed to the excitation of low-energy (spin) degrees of
  freedom in the correlated system.\cite{Vollhardt97} The Maxwell
  relation
  \begin{equation}\label{sv-maxrel}
    \left(\frac{\partial{S}}{\partial{P}}\right)_{\!{T}}
    =-\left(\frac{\partial{V}}{\partial{T}}\right)_{\!{P}}
  \end{equation}
  then implies that at low temperatures
  $\left(\frac{\partial{V}}{\partial{T}}\right)_{\!{P}}<0$. This
  explains the effect discussed above, namely that the volume shrinks
  upon increase of $T$.
  
  As noted before,\cite{Brewer59a,Greywall83} the specific-heat curves
  for different pressures $P$ (or molar volumes $V$) cross sharply at
  $T_+=0.16\mbox{~K}$, a feature which is also observed in other
  strongly correlated systems.\cite{Vollhardt97} In the case of $^3$He
  the crossing of curves is due to the above-mentioned anomalous
  pressure dependence of the specific heat at low temperatures, where
  $\left(\frac{\partial{C_P}}{\partial{P}}\right)_{\!{T}}>0$, and the
  free-fermion behavior
  $\left(\frac{\partial{C_P}}{\partial{P}}\right)_{\!{T}}<0$ at higher
  temperatures (see Fig.~\ref{ct-vs-p}), implying that at some
  intermediate temperature $T_+(P)$ the slope of $C_P$ vs $P$
  vanishes, i.e.,
  $\left(\frac{\partial{C_P}}{\partial{P}}\right)_{\!{T_+}}=0$.
  Consequently, the curves $C_P$ vs $T$ cross at $T_+(P)$ (see
  Fig.~\ref{ct-vs-t}).  Furthermore, the curvature of $C_P$ with
  respect to $P$
  \begin{equation}
    \left(\frac{\partial^2{C_P}}{\partial{P}^2}\right)_{\!{T}}
    =-T
    \left(\frac{\partial^2{}}{\partial{T}^2}\right)_{\!{P}}    
    \left(\frac{\partial{V}}{\partial{P}}\right)_{\!{T}}    
  \end{equation}
  is also small at $T_+$, since
  $\left(\frac{\partial{V}}{\partial{P}}\right)_{\!{T}}=-V\kappa$
  itself is small and depends only weakly on temperature near $T_+$
  (see below).  Therefore at $T_+$ the curves $C_P$ vs  $P$ are
  almost straight lines, implying an essentially {\em pressure
    independent} $T_+(P)$.  The range of temperatures over which the
  crossing occurs is thus very narrow. In fact, the crossing region
  has a width $\Delta C_P/C_P\approx0.5\%$ over the entire pressure
  range,\cite{Vollhardt97} which cannot be resolved on the scale of
  Fig.~\ref{ct-vs-t}, making it appear to be pointlike.

  \subsection{Derivatives of the Molar Volume}\label{subsec:derivs}
  
\begin{figure}[p]
    \centerline{\epsfig{width=\columnwidth,file=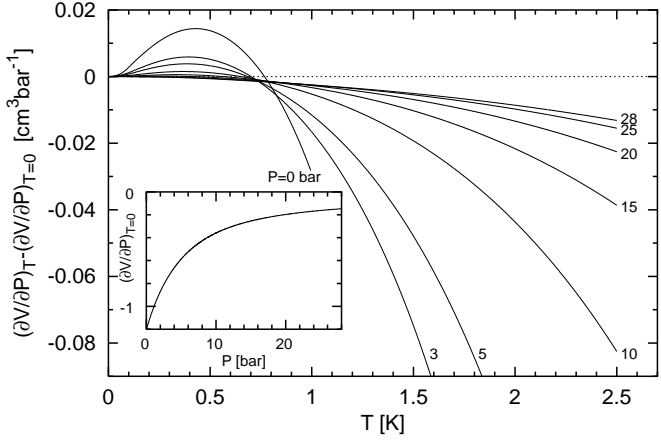}}
    \caption{Change of volume with pressure, 
      $\left(\frac{\partial{V}}{\partial{P}}\right)_{\!{T}}$, vs $T$
      at several pressures, relative to its value at $T=0$
      (inset).\label{dvdp-vs-t-p}}
  \end{figure}

\begin{figure}[p]
    \centerline{\epsfig{width=\columnwidth,file=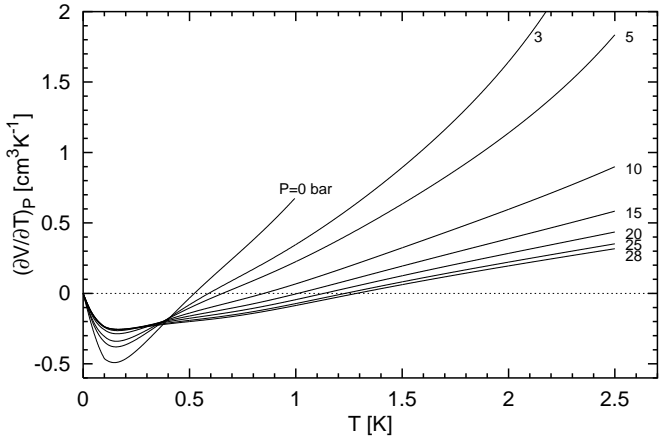}}
    \caption{Change of volume with temperature, 
      $\left(\frac{\partial{V}}{\partial{T}}\right)_{\!{P}}$, vs $T$
      at several pressures $P$.\label{dvdt-vs-t}}
  \end{figure}

  The results for the derivatives of $V$ with respect to $P$ and $T$
  are shown in Figs.~\ref{dvdp-vs-t-p} and \ref{dvdt-vs-t}.  These
  quantities determine the compressibility and the thermal expansion
  coefficient.

  \subsubsection*{Compressibility $\kappa(T,P)$}\label{kappa-subsec}
  
\begin{figure}[p]
    \centerline{\epsfig{width=\columnwidth,file=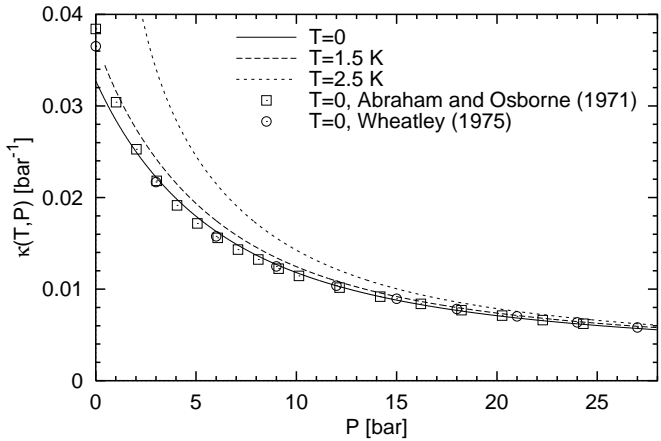}}
    \caption{Compressibility $\kappa(T,P)$ vs $P$
      at temperatures $T=0$, 1.5, 2.5 K, compared to the data at $T=0$
      of Wheatley (Ref.~\protect\onlinecite{Wheatley75}) and Abraham
      and Osborne
      (Ref.~\protect\onlinecite{Abraham71}).\label{k-vs-p}}
  \end{figure}

\begin{figure}[p]
    \centerline{\epsfig{width=\columnwidth,file=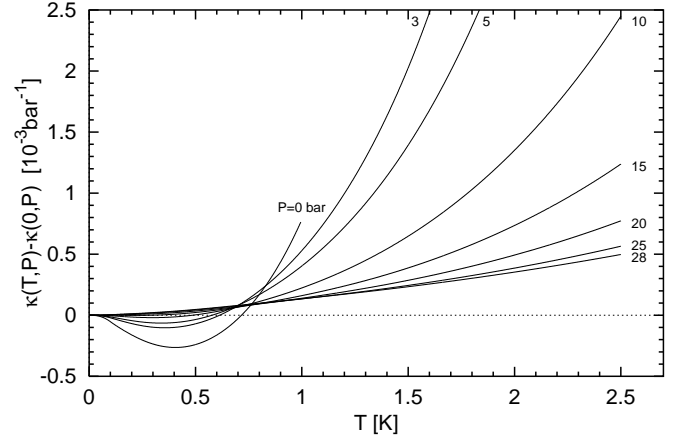}}
    \caption{Compressibility $\kappa(T,P)$ vs $T$
      at several pressures $P$, relative to its value at $T=0$ shown
      in Fig.~\ref{k-vs-p}.\label{k-vs-t}}
  \end{figure}

\begin{figure}[p]
    \centerline{\epsfig{width=\columnwidth,file=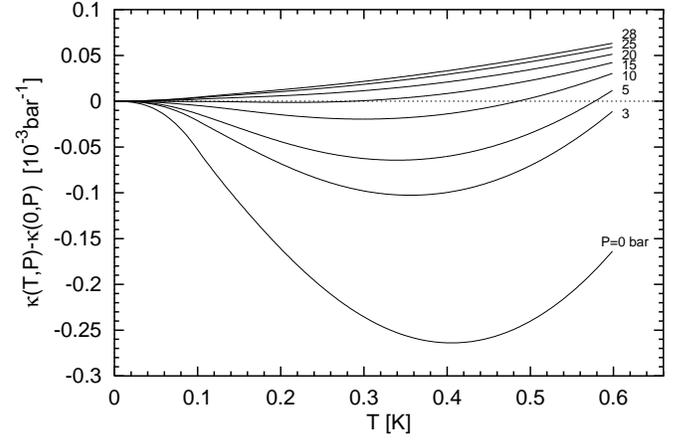}}
    \caption{Low-temperature behavior of the compressibility $\kappa(T,P)$ 
      at several pressures $P$, relative to its value at $T=0$
      (Fig.~\ref{k-vs-p}).\label{k-vs-t-lo}}
  \end{figure}

  We now determine the isothermal compressibility $\kappa(T,P)$
  [Eq.~(\ref{kappadef})] in a wide range of temperatures and
  pressures.  The pressure dependence of $\kappa$ is plotted for
  $T=0,$ 1.5, 2.5 K in Fig.~\ref{k-vs-p}, while Figs.~\ref{k-vs-t} and
  \ref{k-vs-t-lo} show the temperature dependence of $\kappa$ relative
  to its value at $T=0$.  The variation of $\kappa(T,P)$ with
  temperature for a given pressure is rather small, except for high
  temperatures, where the deviation from $\kappa(0,P)$ can become
  quite large (see Fig.~\ref{k-vs-p}).
  
\begin{figure}[p]
    \centerline{\epsfig{width=\columnwidth,file=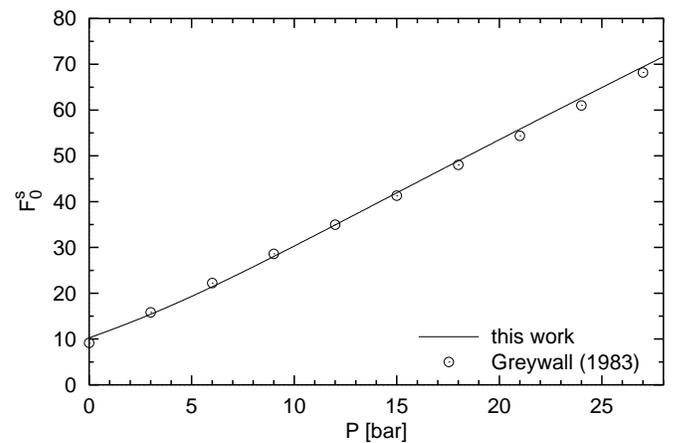}}
    \caption{Fermi liquid parameter $F_0^s$ as a 
      function of pressure, compared to the data obtained by Greywall
      (Ref.~\protect\onlinecite{Greywall83}) using Wheatley's
      (Ref.~\protect\onlinecite{Wheatley75}) values of
      $\kappa(0,P)$.\label{f0-vs-p}}
  \end{figure}

  The zero-temperature value of the compressibility is connected to
  the Landau parameter $F_0^s$ by\cite{Wheatley75}
  \begin{eqnarray}
    \label{kappa0}
    \kappa(0,P)
    &=&\frac{9\pi^2m^*}{\hbar^2(1+F_0^s)}
    \left(\frac{V(0,P)}{3\pi^2N}\right)^{\frac{5}{3}} 
    \nonumber\\
    &=&3.285\cdot10^{-4}\,\mbox{bar}^{-1}
    \frac{m^*/m}{1+F_0^s}
    \left(\frac{V(0,P)}{\mbox{cm}^3}\right)^\frac{5}{3}
    .
  \end{eqnarray}
  Using our evaluation of $\kappa$ we can calculate $F_0^s$ as a
  function of $P$, as shown in Fig.~\ref{f0-vs-p}.
  Greywall's\cite{Greywall83} data, obtained by adjusting Wheatley's
  values for $F_0^s$ according to his new determination of $m^*/m$,
  are also shown. They are in good agreement with our evaluation.
  
  We can also  compare our evaluation of $\kappa(0,P)$ to that
  obtained from measurements of the sound velocity and molar volume as
  listed by Wheatley,\cite{Wheatley75} as well as to the data
  published by Abraham and Osborne\cite{Abraham71} which they obtained
  by differentiating the molar volume. These data are included in
  Fig.~\ref{k-vs-p}. They agree well with our evaluation, with
  relative differences of at most 2\%, except at  very low pressures
  ($P$ $\lesssim$ 2~bar).  At higher temperatures our evaluation of
  $\kappa(T,P)$ can again be compared to Abraham and
  Osborne's\cite{Abraham71} data, where we also find good agreement
  except for $P\lesssim2~\mbox{bar}$.  There is also good agreement
  with the data for $\kappa(T,P)$ published by Grilly.\cite{Grilly71}
  
  The difference at low pressures between our results for
  $\kappa(0,P)$ and other data, e.g.\ by Wheatley,\cite{Wheatley75}
  does not imply that any of Greywall's original data for $C_V$ or his
  formula for $V(0.1\mbox{K},P)$ are inaccurate.  It merely indicates
  that our calculation of the {\em derivative}
  $\left(\frac{\partial{V}}{\partial{P}}\right)_{\!{T}}$ based on
  Greywall's interpolation formula for $V(T_0,P)$ may not be
  sufficently accurate at low pressures.\cite{footnote3} Therefore the
  results for $\kappa(T,P)$ may indeed underestimate the pressure
  dependence of the molar volume at $P\lesssim2~\mbox{bar}$.

  \subsubsection*{Thermal expansion coefficient $\alpha(T,P)$} 
  
\begin{figure}[p]
    \centerline{\epsfig{width=\columnwidth,file=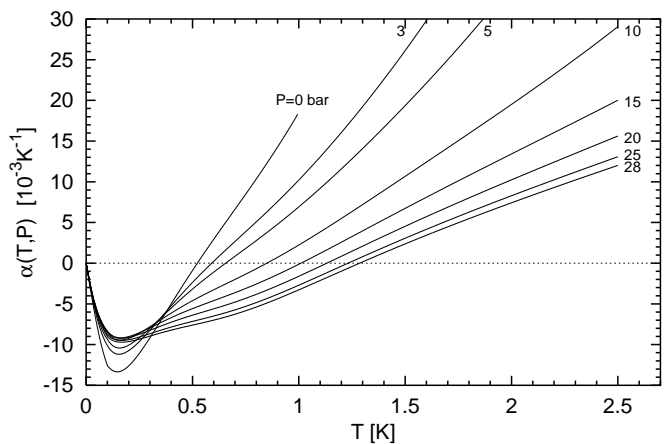}}
    \caption{Thermal expansion coefficient $\alpha(T,P)$ 
      vs $T$ at several pressures $P$.\label{a-vs-t-hi}}
  \end{figure}

\begin{figure}[p]
    \centerline{\epsfig{width=\columnwidth,file=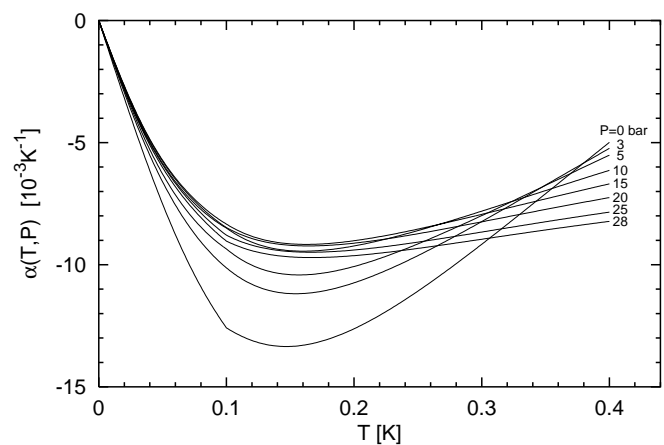}}%
    \caption{Low-temperature behavior of
      the thermal expansion coefficient $\alpha(T,P)$ at several
      pressures $P$.\label{a-vs-t-lo}}
  \end{figure}

\begin{figure}[p]
    \centerline{\epsfig{width=\columnwidth,file=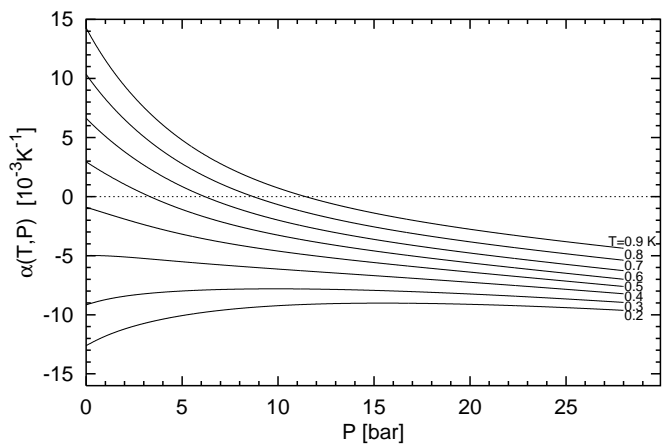}}
    \caption{Thermal expansion coefficient $\alpha(T,P)$ 
      vs $P$ at several temperatures $T$.\label{a-vs-p}}
  \end{figure}

\begin{figure}[p]
    \centerline{\epsfig{width=\columnwidth,file=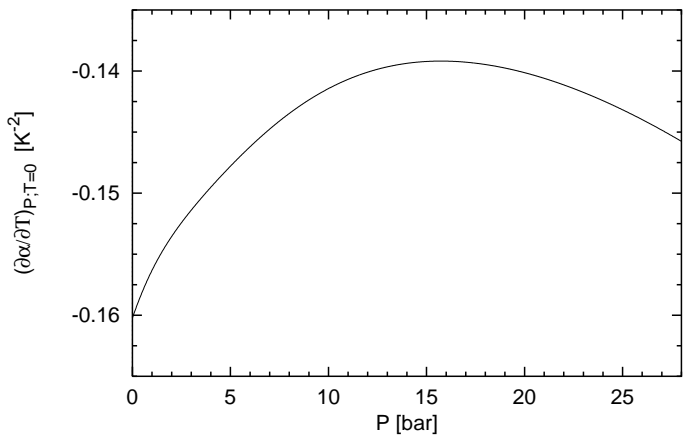}}
    \caption{Pressure dependence
      of the thermal expansion coefficient,
      $\left(\frac{\partial{\alpha}}{\partial{T}}\right)_{\!{P}}$, in
      the limit $T\to0$.\label{dadt-vs-p}}
  \end{figure}

  The temperature dependence of $\alpha(T,P)$ [Eq.~(\ref{alphadef})]
  is presented in Fig.~\ref{a-vs-t-hi}, with Fig.~\ref{a-vs-t-lo}
  showing the low-temperature region in greater detail. The pressure
  dependence is plotted in Fig.~\ref{a-vs-p}.\cite{footnote5}  Using
  Eq.~(\ref{sv-maxrel}) and the pressure dependence of $\gamma(P)$ the
  expansion coefficient is seen to be linear in $T$ at very low
  temperatures,
  \begin{equation}
    \alpha(T,P)=-\frac{\gamma'(P)}{V(0,P)}\,T,\;\;\;T\to0,
  \end{equation}
  i.e., is {\em negative} and vanishes at $T=0$.  The slope
  $\left(\frac{\partial{\alpha}}{\partial{T}}\right)_{\!{P;T=0}}=
  -\gamma'(P)/V(0,P)<0$ is plotted in Fig.~\ref{dadt-vs-p} as a
  function of pressure. At low pressures, it increases upon increase
  of pressure, but decreases again for pressures $P$ $\gtrsim$ 15~bar.
  This surprising nonmonotonic behavior implies that the curves of
  $\alpha$ cross at very low temperatures.
  
\begin{figure}[p]
    \centerline{\epsfig{width=\columnwidth,file=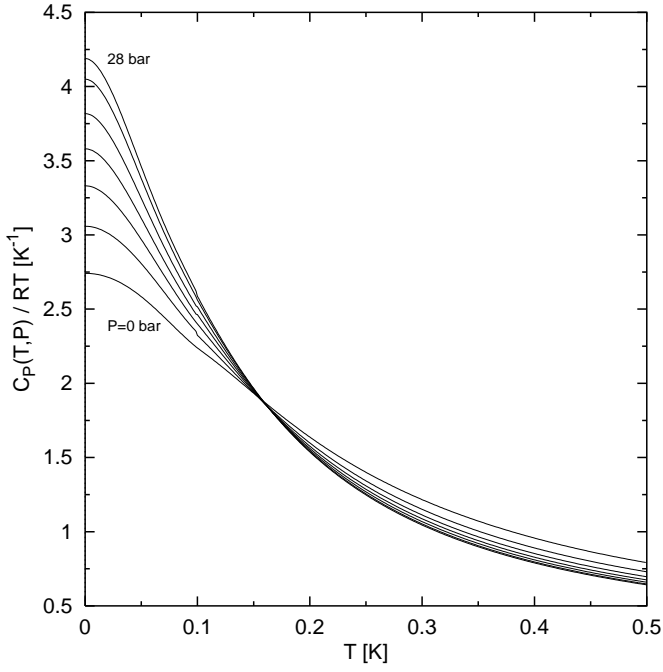}}
    \caption{Specific heat at
      constant pressure $C_P(T,P)$ divided by temperature vs $T$ at
      pressures $P=0,$ 5, 10, 15, 20, 25, 28 bar. The small kinks at
      the temperature $T_0=0.1\mbox{~K}$ in this and other figures are
      artefacts caused by the different interpolation formulas
      (Ref.~\protect\onlinecite{Greywall83}) for $C_V(T,V)$ used below
      and above $T_0$.\label{ct-vs-t}}
  \end{figure}

  At higher temperatures the expansion coefficient shows ``normal''
  behavior with $\alpha>0$, leading to a minimum in $\alpha$ vs $T$.
  Since
  $\alpha^2\ll\left(\frac{\partial{\alpha}}{\partial{T}}\right)_{\!{P}}$
  at low temperatures, the thermodynamic relation
  \begin{equation}
    \left(\frac{\partial{C_P}}{\partial{P}}\right)_{\!{T}}    
    =-VT\left[\alpha^2+
      \left(\frac{\partial{\alpha}}{\partial{T}}\right)_{\!{P}}
    \right]
  \end{equation}
  implies $\left(\frac{\partial{\alpha}}{\partial{T}}\right)_{\!{P}}$
  $\approx$
  $-\frac{1}{VT}\left(\frac{\partial{C_P}}{\partial{P}}\right)_{\!{T}}$
  $=$ $0$ at $T_+=0.16\mbox{~K}$, i.e., the minima of $\alpha$ are
  essentially all located at $T_+$ where the specific-heat curves
  cross (see Figs.~\ref{a-vs-t-hi},\ref{a-vs-t-lo}).  The very weak
  pressure dependence of these minima is due to the small width of the
  crossing region of the specific-heat curves (see Fig.~\ref{ct-vs-t})
  as discussed above.
  
  Our evaluation may be compared with earlier determinations of
  $\alpha(T,P)$. We find good agreement with the data of Abraham and
  Osborne;\cite{Abraham71} for $P\gtrsim1\mbox{~bar}$ the differences
  are below $1.6\cdot10^{-3}\mbox{K}^{-1}$ except for
  $0.1 \leq T\leq0.4\mbox{~K}$, where they are as high as
  $10^{-2}\mbox{K}^{-1}$ due to different locations of the minima of
  $\alpha(T,P)$. Our data also agree with the data by Boghosian {\em
    et al.}\cite{Boghosian66} listed in Wilks' book\cite{Wilks67}
  within $10^{-3}\mbox{K}^{-1}$ for $T<1\mbox{~K}$ or
  $P>10\mbox{~bar}$, and $6\cdot10^{-3}\mbox{K}^{-1}$ for
  $T\geq1\mbox{~K}$ and $P\lesssim10\mbox{~bar}$. However, there are
  also qualitative differences, as we will now discuss.
  
  In Fig.~\ref{a-vs-t-lo} the curves of $\alpha(T,P)$ are seen to
  cross near $T\approx0.35~\mbox{K}$ in a rather broad range of
  temperatures, as was already observed by Lee {\em et
    al.}\cite{Lee61} By contrast, in some of the earlier measurements
  of $\alpha(T,P)$ a {\em sharp} crossing of these curves had been
  reported.  For example, Brewer and Daunt\cite{Brewer59b} presented a
  figure with a sharp crossing point of the curves at $T=0.4\mbox{~K}$
  for pressures in the range from 2 to 22 bar, but noted that the
  sharpness of this feature could not be decided with certainty.  In
  Wilks' book\cite{Wilks67} $\alpha(T,P)$ is also plotted with a sharp
  crossing point at $T=0.35\mbox{~K}$ for pressures 0-25~bar,
  although the error bars are of the order of $10^{-3}\mbox{K}^{-1}$,
  i.e., are much larger than the suggested accuracy of the crossing
  point.
  
\begin{figure}[p]
    \centerline{\epsfig{width=\columnwidth,file=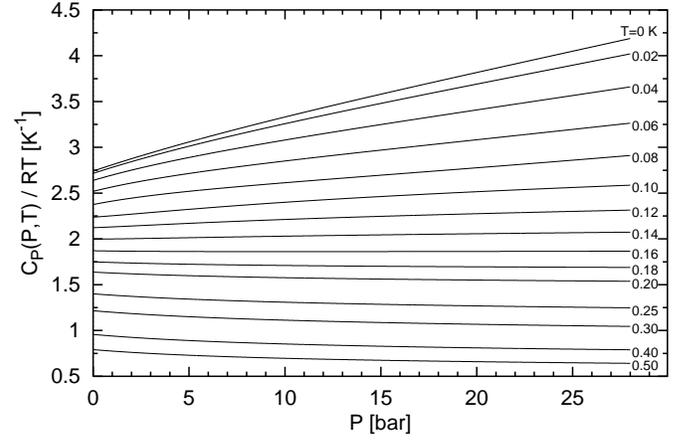}}
    \caption{Specific heat at
      constant pressure $C_P(T,P)$ divided by temperature vs $P$ at
      several temperatures $T$. At $T_+=0.16\mbox{~K}$, $C_P(T,P)$ is
      essentially independent of pressure over the whole pressure
      range, leading to a sharp crossing point in the specific heat
      (see text).\label{ct-vs-p}}
  \end{figure}

  Our evaluation of the expansion coefficient clearly shows that the
  curves of $\alpha(T,P)$ do not cross at a single well-defined
  temperature.  The condition for a crossing at some temperature
  $T'_+(P)$ is
  $\left(\frac{\partial{\alpha}}{\partial{P}}\right)_{\!{T}}=0$; for
  $P$ $=$ 10 bar we have $T'_{+}\approx0.3\mbox{~K}$.  From the
  crossing condition we can estimate the width of the crossing region
  as
  \begin{equation}
    T'_+(P)=T'_{+}(10\mbox{bar})
    +\left[
      \left(\frac{\partial{\alpha}}{\partial{P}}\right)_{\!{T}}      
      \Big/
      \left(\frac{\partial^2{\alpha}}{\partial{P}\partial{T}}\right)
    \right]_{T'_{+}(10\text{bar})}.
  \end{equation}
  In the pressure range $5 \leq P\leq25\mbox{~bar}$ the
  numerator of the second term on the right hand side is of the order
  of $10^{-4}\mbox{K}^{-1}\mbox{bar}^{-1}$, while the denominator is
  approximately $10^{-5}\mbox{K}^{-2}\mbox{bar}^{-1}$. These values
  can be determined from Fig.~\ref{a-vs-p} and Tables
  \ref{firsttable}-\ref{lasttable}, and are also consistent with the
  data of Refs.~\onlinecite{Wilks67}.  Hence $T'_+$ varies by
  approximately $0.1\mbox{~K}$ over this pressure range, which is also
  consistent with the observed width of the crossing region in
  Fig.~\ref{a-vs-t-lo}. This pressure dependence of $T'_+$ can also be
  recognized from Fig.~\ref{a-vs-p} as the fact that $\alpha(T,P)$
  never becomes a horizontal line as a function of $P$, contrary to
  the situation for the specific heat (Fig.~\ref{ct-vs-p}).  We
  conclude that for the curves of the expansion coefficient there
  exists only a broad crossing region, i.e., no sharp crossing point,
  in contrast to the data presented by Wilks.\cite{Wilks67}
  
  In summary, we showed that Greywall's\cite{Greywall83} data contain
  the complete thermodynamic information for the normal-liquid phase
  of $^3$He. This enabled us to calculate all equilibrium
  thermodynamic quantities for this phase at zero magnetic field for a
  wide range of parameters.  We plotted and tabulated these data as
  reference material and benchmark for future investigations.  In
  general, the published experimental data agree well with our
  evaluations.  However the sharp crossing point of the curves of the
  expansion coefficient vs temperature for different pressures,
  summarized in Ref.~\onlinecite{Wilks67}, was found to be an artefact
  since it is not found in the thermodynamically consistent data
  presented here.  The difference in the compressibility at low
  pressures ($P$ $\lesssim$ 2~bar) between our values and the data
  obtained from Ref.~\onlinecite{Wheatley75} makes it highly desirable
  to have available high-precision data of the molar volume at a
  single temperature, which could be used in place of our
  Eq.~(\ref{p0def}). This would lead, in principle, to further
  improvements in the accuracy of the thermodynamic quantities
  calculated in this paper.  The data of this paper may be calculated
  interactively for arbitrary input parameters on our
  website.\cite{www}
  
  \emph{Note added in proof}: After our paper went into production the
  experimental work by P. R. Roach \emph{et al.} [J. Low Temp. Phys.
  \textbf{52}, 433 (1983)] came to our attention, where the isobaric
  expansion coefficient and the adiabatic compressibility were
  measured.  Comparing their results for $P(T,V)-P(0,V)$ at three
  different temperatures with those obtained by Greywall (Ref. 16),
  the authors find a discrepancy of about $0.1$ bar at $T=0.6$ K and
  high pressures which they attribute to Greywall's interpolation
  formulae, Eqs. (7)-(9).

  \cleardoublepage

  \clearpage

  \begin{table*}[p]
    \begin{ruledtabular}\begin{tabular}{l|r@{}l|r@{}l|r@{}l|r@{}l}
      &\multicolumn{2}{c|}{$j=0$}&
      \multicolumn{2}{c|}{$j=1$}&
      \multicolumn{2}{c|}{$j=2$}&
      \multicolumn{2}{c}{$j=3$}\\
      \hline
      $a_{1j}$ &
      -2.9190414& &
      5.2893401&${{\cdot}}10^{2}$&
      -1.8869641&${{\cdot}}10^{4}$&
      2.6031315&${{\cdot}}10^{5}$
      \\
      $a_{3j}$ &
      -2.4752597&${{\cdot}}10^{3}$& 
      1.8377260&${{\cdot}}10^{5}$&
      -3.4946553&${{\cdot}}10^{6}$&
      0\phantom{.0000000}
      \\
      $a_{4j}$ &
      3.8887481&${{\cdot}}10^{4}$&
      -2.8649769&${{\cdot}}10^{6}$&
      5.2526785&${{\cdot}}10^{7}$&
      0\phantom{.0000000}
      \\
      $a_{5j}$ &
      -1.7505655&${{\cdot}}10^{5}$&
      1.2809001&${{\cdot}}10^{7}$&
      -2.3037701&${{\cdot}}10^{8}$&
      0\phantom{.0000000}
      \\
      \hline
      $b_{0j}$ &  
      -6.5521193&${{\cdot}}10^{-2}$& 
      1.3502371&${{\cdot}}10^{-2}$& 
      0\phantom{.0000000} & &
      0\phantom{.0000000}
      \\
      $b_{1j}$ &
      4.1359033&${{\cdot}}10^{-2}$&
      3.8233755&${{\cdot}}10^{-4}$&
      -5.3468396&${{\cdot}}10^{-5}$&
      0\phantom{.0000000}
      \\
      $b_{2j}$ &
      5.7976786&${{\cdot}}10^{-3}$&
      -6.5611532&${{\cdot}}10^{-4}$&
      1.2689707&${{\cdot}}10^{-5}$&
      0\phantom{.0000000}
      \\
      $b_{3j}$ &
      -3.8374623&${{\cdot}}10^{-4}$&
      3.2072581&${{\cdot}}10^{-5}$&
      -5.3038906&${{\cdot}}10^{-7}$&
      0\phantom{.0000000}
      \\
      \hline
      $c_{1j}$ &
      -2.5482958&${{\cdot}}10^{1}$&
      1.6416936&  & -1.5110378&
      ${{\cdot}}10^{-2}$&
      0\phantom{.0000000}
      \\
      $c_{2j}$ &
      3.7882751&${{\cdot}}10^{1}$&
      -2.8769188&  &
      3.5751181&${{\cdot}}10^{-2}$&
      0\phantom{.0000000}
      \\
      $c_{3j}$ &
      2.4412956&${{\cdot}}10^{1}$&
      -2.4244083&  &
      6.7775905&${{\cdot}}10^{-2}$&
      0\phantom{.0000000}
      \\
      \hline
      $d_{j}$ &
      -7.1613436&&
      6.0525139&${{\cdot}}10^{-1}$&
      -7.1295855&${{\cdot}}10^{-3}$&
      0\phantom{.0000000}
      \\
      \hline
      $a_{j}$&
      3.6820\phantom{000}&${{\cdot}}10^{1}$&
      -1.2094\phantom{000}& &
      9.4231\phantom{000}&${{\cdot}}10^{-2}$&
      -4.9875\phantom{000}&${{\cdot}}10^{-3}$
      \\
      $a_{j+4}$&1.3746\phantom{000}&${{\cdot}}10^{-4}$&
      -1.4756\phantom{000}&${{\cdot}}10^{-6}$&
      0\phantom{.0000000}& &
      0\phantom{.0000000}
      \\
      \hline
      $b_{j}$&
      0\phantom{.0000000}& &
      -8.3094892&${{\cdot}}10^{-1}$&
      6.1583050&${{\cdot}}10^{-2}$&
      -4.5946040&${{\cdot}}10^{-3}$
      \\
      $b_{j+4}$&
      1.7370990&${{\cdot}}10^{-4}$&
      -3.8137958&${{\cdot}}10^{-5}$&
      2.3397112&${{\cdot}}10^{-6}$&
      1.7579799&${{\cdot}}10^{-7}$
    \end{tabular}\end{ruledtabular}
    \caption{Coefficients entering the interpolation
      formulas (Ref.~\protect\onlinecite{Greywall83})
      for the specific heat
      [Eqs.~(\ref{c1def}) and (\ref{c2def})],
      the molar volume  [Eq.~(\ref{v0def})],
      and the least-square fit for the pressure 
      [Eq.~(\ref{p0def})].\label{paramtable}}
  \end{table*}

\begin{table*}[p]
\begin{ruledtabular}\begin{tabular}{ddddddd}
\multicolumn{1}{r}{$T$} &
\multicolumn{1}{r}{$S/R$} &
\multicolumn{1}{r}{$C_P/RT$} &
\multicolumn{1}{r}{$V$} &
\multicolumn{1}{r}{$-(\partial V/\partial P)_T$} &
\multicolumn{1}{r}{$\kappa$} &
\multicolumn{1}{r}{$\alpha$}\\
\multicolumn{1}{r}{$[\mbox{K}]$} &
\multicolumn{1}{r}{} &
\multicolumn{1}{r}{$[\mbox{K}^{-1}]$} &
\multicolumn{1}{r}{$[\mbox{cm}^3]$} &
\multicolumn{1}{r}{$[\mbox{cm}^3\mbox{bar}^{-1}]$} &
\multicolumn{1}{r}{$[10^{-2}\mbox{bar}^{-1}]$} &
\multicolumn{1}{r}{$[10^{-3}\mbox{K}^{-1}]$}\\
\hline
0     & 0      & 2.7411 & 36.8461 & 1.2062 & 3.27375 &    0\\
0.005 & 0.0137 & 2.7395 & 36.8460 & 1.2062 & 3.27374 &   -0.80\\
0.010 & 0.0274 & 2.7348 & 36.8458 & 1.2062 & 3.27372 &   -1.60\\
0.020 & 0.0547 & 2.7156 & 36.8449 & 1.2062 & 3.27364 &   -3.17\\
0.030 & 0.0817 & 2.6835 & 36.8435 & 1.2061 & 3.27348 &   -4.70\\
0.040 & 0.1083 & 2.6390 & 36.8415 & 1.2059 & 3.27322 &   -6.15\\
0.060 & 0.1599 & 2.5193 & 36.8360 & 1.2054 & 3.27228 &   -8.77\\
0.080 & 0.2089 & 2.3753 & 36.8287 & 1.2046 & 3.27070 &   -10.91\\
0.100 & 0.2550 & 2.2376 & 36.8200 & 1.2034 & 3.26845 &   -12.59\\
0.150 & 0.3595 & 1.9337 & 36.7959 & 1.2005 & 3.26267 &   -13.35\\
0.200 & 0.4485 & 1.6378 & 36.7718 & 1.1979 & 3.25765 &   -12.62\\
0.250 & 0.5242 & 1.4004 & 36.7499 & 1.1956 & 3.25341 &   -11.08\\
0.300 & 0.5895 & 1.2160 & 36.7313 & 1.1938 & 3.25020 &   -9.16\\
0.350 & 0.6465 & 1.0716 & 36.7164 & 1.1926 & 3.24815 &   -7.09\\
0.400 & 0.6971 & 0.9570 & 36.7053 & 1.1920 & 3.24736 &   -4.99\\
0.450 & 0.7426 & 0.8652 & 36.6981 & 1.1919 & 3.24788 &   -2.91\\
0.500 & 0.7839 & 0.7913 & 36.6946 & 1.1925 & 3.24973 &   -0.90\\
0.600 & 0.8573 & 0.6834 & 36.6984 & 1.1955 & 3.25751 &    2.94\\
0.700 & 0.9219 & 0.6129 & 36.7160 & 1.2010 & 3.27100 &    6.63\\
0.800 & 0.9807 & 0.5661 & 36.7471 & 1.2092 & 3.29069 &    10.35\\
0.900 & 1.0356 & 0.5343 & 36.7923 & 1.2205 & 3.31734 &    14.27\\
\end{tabular}\end{ruledtabular}
\caption{Thermodynamic functions for normal-liquid
  $^3$He at pressure $P=0\,\mbox{bar}$.
  See also the discussion of the accuracy of
  $\left(\frac{\partial{V}}{\partial{T}}\right)_{\!{P}}$ 
  at low pressures in Sec.~IV.C and
  Ref.~\protect\onlinecite{footnote3}.\label{firsttable}}
\end{table*}

\begin{table*}[p]
\begin{ruledtabular}\begin{tabular}{ddddddd}
\multicolumn{1}{r}{$T$} &
\multicolumn{1}{r}{$S/R$} &
\multicolumn{1}{r}{$C_P/RT$} &
\multicolumn{1}{r}{$V$} &
\multicolumn{1}{r}{$-(\partial V/\partial P)_T$} &
\multicolumn{1}{r}{$\kappa$} &
\multicolumn{1}{r}{$\alpha$}\\
\multicolumn{1}{r}{$[\mbox{K}]$} &
\multicolumn{1}{r}{} &
\multicolumn{1}{r}{$[\mbox{K}^{-1}]$} &
\multicolumn{1}{r}{$[\mbox{cm}^3]$} &
\multicolumn{1}{r}{$[\mbox{cm}^3\mbox{bar}^{-1}]$} &
\multicolumn{1}{r}{$[10^{-2}\mbox{bar}^{-1}]$} &
\multicolumn{1}{r}{$[10^{-3}\mbox{K}^{-1}]$}\\
\hline
0     & 0      & 3.0586 & 32.6559 & 0.5863 & 1.79539 &    0\\
0.005 & 0.0153 & 3.0555 & 32.6558 & 0.5863 & 1.79539 &   -0.74\\
0.010 & 0.0305 & 3.0466 & 32.6556 & 0.5863 & 1.79539 &   -1.47\\
0.020 & 0.0609 & 3.0124 & 32.6549 & 0.5863 & 1.79536 &   -2.86\\
0.030 & 0.0907 & 2.9589 & 32.6538 & 0.5862 & 1.79531 &   -4.14\\
0.040 & 0.1200 & 2.8891 & 32.6522 & 0.5862 & 1.79523 &   -5.29\\
0.060 & 0.1761 & 2.7147 & 32.6481 & 0.5860 & 1.79496 &   -7.14\\
0.080 & 0.2284 & 2.5199 & 32.6430 & 0.5858 & 1.79455 &   -8.46\\
0.100 & 0.2770 & 2.3232 & 32.6372 & 0.5855 & 1.79401 &   -9.35\\
0.150 & 0.3835 & 1.9367 & 32.6208 & 0.5847 & 1.79246 &   -10.41\\
0.200 & 0.4715 & 1.5987 & 32.6040 & 0.5839 & 1.79098 &   -10.07\\
0.250 & 0.5448 & 1.3426 & 32.5883 & 0.5833 & 1.78982 &   -9.15\\
0.300 & 0.6069 & 1.1507 & 32.5743 & 0.5828 & 1.78913 &   -7.99\\
0.350 & 0.6606 & 1.0043 & 32.5623 & 0.5825 & 1.78897 &   -6.75\\
0.400 & 0.7078 & 0.8905 & 32.5523 & 0.5825 & 1.78937 &   -5.51\\
0.450 & 0.7500 & 0.8009 & 32.5443 & 0.5827 & 1.79035 &   -4.32\\
0.500 & 0.7882 & 0.7298 & 32.5382 & 0.5831 & 1.79189 &   -3.18\\
0.600 & 0.8557 & 0.6268 & 32.5313 & 0.5845 & 1.79665 &   -1.09\\
0.700 & 0.9147 & 0.5582 & 32.5310 & 0.5867 & 1.80350 &    0.85\\
0.800 & 0.9680 & 0.5102 & 32.5369 & 0.5897 & 1.81231 &    2.78\\
0.900 & 1.0172 & 0.4749 & 32.5491 & 0.5934 & 1.82304 &    4.79\\
1.000 & 1.0633 & 0.4481 & 32.5681 & 0.5979 & 1.83573 &    6.92\\
1.200 & 1.1489 & 0.4112 & 32.6282 & 0.6094 & 1.86756 &    11.58\\
1.400 & 1.2287 & 0.3892 & 32.7203 & 0.6248 & 1.90941 &    16.68\\
1.600 & 1.3052 & 0.3767 & 32.8474 & 0.6449 & 1.96331 &    22.13\\
1.800 & 1.3798 & 0.3705 & 33.0121 & 0.6707 & 2.03183 &    27.95\\
2.000 & 1.4537 & 0.3687 & 33.2178 & 0.7037 & 2.11835 &    34.28\\
2.500 & 1.6398 & 0.3794 & 33.9491 & 0.8327 & 2.45273 &    54.08\\
\end{tabular}\end{ruledtabular}
\caption{Thermodynamic functions for normal-liquid
  $^3$He at pressure $P=5\,\mbox{bar}$.}
\end{table*}

\begin{table*}[p]
\begin{ruledtabular}\begin{tabular}{ddddddd}
\multicolumn{1}{r}{$T$} &
\multicolumn{1}{r}{$S/R$} &
\multicolumn{1}{r}{$C_P/RT$} &
\multicolumn{1}{r}{$V$} &
\multicolumn{1}{r}{$-(\partial V/\partial P)_T$} &
\multicolumn{1}{r}{$\kappa$} &
\multicolumn{1}{r}{$\alpha$}\\
\multicolumn{1}{r}{$[\mbox{K}]$} &
\multicolumn{1}{r}{} &
\multicolumn{1}{r}{$[\mbox{K}^{-1}]$} &
\multicolumn{1}{r}{$[\mbox{cm}^3]$} &
\multicolumn{1}{r}{$[\mbox{cm}^3\mbox{bar}^{-1}]$} &
\multicolumn{1}{r}{$[10^{-2}\mbox{bar}^{-1}]$} &
\multicolumn{1}{r}{$[10^{-3}\mbox{K}^{-1}]$}\\
\hline
0     & 0      & 3.3313 & 30.3745 & 0.3587 & 1.18080 &    0\\
0.005 & 0.0166 & 3.3261 & 30.3744 & 0.3587 & 1.18080 &   -0.71\\
0.010 & 0.0332 & 3.3115 & 30.3743 & 0.3587 & 1.18079 &   -1.40\\
0.020 & 0.0661 & 3.2580 & 30.3737 & 0.3586 & 1.18078 &   -2.72\\
0.030 & 0.0983 & 3.1784 & 30.3726 & 0.3586 & 1.18076 &   -3.92\\
0.040 & 0.1296 & 3.0799 & 30.3713 & 0.3586 & 1.18072 &   -4.96\\
0.060 & 0.1890 & 2.8510 & 30.3678 & 0.3585 & 1.18062 &   -6.60\\
0.080 & 0.2436 & 2.6136 & 30.3634 & 0.3584 & 1.18049 &   -7.74\\
0.100 & 0.2937 & 2.4003 & 30.3584 & 0.3583 & 1.18033 &   -8.50\\
0.150 & 0.4021 & 1.9436 & 30.3446 & 0.3580 & 1.17983 &   -9.42\\
0.200 & 0.4896 & 1.5765 & 30.3304 & 0.3577 & 1.17932 &   -9.23\\
0.250 & 0.5614 & 1.3089 & 30.3168 & 0.3574 & 1.17897 &   -8.60\\
0.300 & 0.6217 & 1.1127 & 30.3044 & 0.3572 & 1.17884 &   -7.81\\
0.350 & 0.6735 & 0.9656 & 30.2932 & 0.3572 & 1.17899 &   -6.96\\
0.400 & 0.7188 & 0.8529 & 30.2833 & 0.3572 & 1.17942 &   -6.13\\
0.450 & 0.7592 & 0.7653 & 30.2746 & 0.3573 & 1.18013 &   -5.35\\
0.500 & 0.7957 & 0.6964 & 30.2671 & 0.3575 & 1.18112 &   -4.61\\
0.600 & 0.8600 & 0.5967 & 30.2552 & 0.3582 & 1.18388 &   -3.26\\
0.700 & 0.9161 & 0.5290 & 30.2472 & 0.3592 & 1.18756 &   -1.99\\
0.800 & 0.9664 & 0.4799 & 30.2432 & 0.3605 & 1.19204 &   -0.68\\
0.900 & 1.0124 & 0.4426 & 30.2432 & 0.3621 & 1.19724 &    0.72\\
1.000 & 1.0552 & 0.4136 & 30.2477 & 0.3639 & 1.20316 &    2.23\\
1.200 & 1.1335 & 0.3732 & 30.2709 & 0.3685 & 1.21735 &    5.50\\
1.400 & 1.2055 & 0.3489 & 30.3146 & 0.3744 & 1.23512 &    8.93\\
1.600 & 1.2738 & 0.3347 & 30.3793 & 0.3819 & 1.25707 &    12.42\\
1.800 & 1.3398 & 0.3263 & 30.4656 & 0.3911 & 1.28378 &    15.94\\
2.000 & 1.4045 & 0.3213 & 30.5737 & 0.4023 & 1.31592 &    19.51\\
2.500 & 1.5636 & 0.3162 & 30.9456 & 0.4412 & 1.42576 &    29.06\\
\end{tabular}\end{ruledtabular}
\caption{Thermodynamic functions for normal-liquid
  $^3$He at pressure $P=10\,\mbox{bar}$.}
\end{table*}

\begin{table*}[p]
\begin{ruledtabular}\begin{tabular}{ddddddd}
\multicolumn{1}{r}{$T$} &
\multicolumn{1}{r}{$S/R$} &
\multicolumn{1}{r}{$C_P/RT$} &
\multicolumn{1}{r}{$V$} &
\multicolumn{1}{r}{$-(\partial V/\partial P)_T$} &
\multicolumn{1}{r}{$\kappa$} &
\multicolumn{1}{r}{$\alpha$}\\
\multicolumn{1}{r}{$[\mbox{K}]$} &
\multicolumn{1}{r}{} &
\multicolumn{1}{r}{$[\mbox{K}^{-1}]$} &
\multicolumn{1}{r}{$[\mbox{cm}^3]$} &
\multicolumn{1}{r}{$[\mbox{cm}^3\mbox{bar}^{-1}]$} &
\multicolumn{1}{r}{$[10^{-2}\mbox{bar}^{-1}]$} &
\multicolumn{1}{r}{$[10^{-3}\mbox{K}^{-1}]$}\\
\hline
0     & 0      & 3.5803 & 28.8722 & 0.2540 & 0.87982 &    0\\
0.005 & 0.0179 & 3.5731 & 28.8721 & 0.2540 & 0.87982 &   -0.69\\
0.010 & 0.0357 & 3.5528 & 28.8720 & 0.2540 & 0.87982 &   -1.38\\
0.020 & 0.0709 & 3.4799 & 28.8714 & 0.2540 & 0.87982 &   -2.67\\
0.030 & 0.1052 & 3.3747 & 28.8704 & 0.2540 & 0.87982 &   -3.84\\
0.040 & 0.1383 & 3.2487 & 28.8692 & 0.2540 & 0.87981 &   -4.86\\
0.060 & 0.2005 & 2.9696 & 28.8659 & 0.2540 & 0.87980 &   -6.45\\
0.080 & 0.2572 & 2.6959 & 28.8618 & 0.2539 & 0.87979 &   -7.56\\
0.100 & 0.3086 & 2.4629 & 28.8572 & 0.2539 & 0.87979 &   -8.33\\
0.150 & 0.4185 & 1.9504 & 28.8444 & 0.2538 & 0.87975 &   -9.14\\
0.200 & 0.5058 & 1.5617 & 28.8313 & 0.2536 & 0.87969 &   -9.02\\
0.250 & 0.5766 & 1.2859 & 28.8186 & 0.2535 & 0.87972 &   -8.54\\
0.300 & 0.6356 & 1.0870 & 28.8067 & 0.2535 & 0.87988 &   -7.94\\
0.350 & 0.6861 & 0.9397 & 28.7958 & 0.2535 & 0.88019 &   -7.30\\
0.400 & 0.7302 & 0.8282 & 28.7857 & 0.2535 & 0.88066 &   -6.69\\
0.450 & 0.7694 & 0.7423 & 28.7765 & 0.2536 & 0.88129 &   -6.11\\
0.500 & 0.8047 & 0.6749 & 28.7681 & 0.2538 & 0.88208 &   -5.58\\
0.600 & 0.8670 & 0.5771 & 28.7535 & 0.2542 & 0.88407 &   -4.60\\
0.700 & 0.9211 & 0.5091 & 28.7416 & 0.2548 & 0.88652 &   -3.62\\
0.800 & 0.9694 & 0.4584 & 28.7327 & 0.2555 & 0.88934 &   -2.56\\
0.900 & 1.0132 & 0.4192 & 28.7270 & 0.2564 & 0.89246 &   -1.39\\
1.000 & 1.0535 & 0.3885 & 28.7248 & 0.2573 & 0.89588 &   -0.13\\
1.200 & 1.1266 & 0.3463 & 28.7318 & 0.2597 & 0.90374 &    2.58\\
1.400 & 1.1932 & 0.3216 & 28.7546 & 0.2626 & 0.91318 &    5.36\\
1.600 & 1.2559 & 0.3073 & 28.7933 & 0.2662 & 0.92445 &    8.10\\
1.800 & 1.3165 & 0.2989 & 28.8478 & 0.2705 & 0.93777 &    10.78\\
2.000 & 1.3757 & 0.2936 & 28.9177 & 0.2757 & 0.95333 &    13.42\\
2.500 & 1.5203 & 0.2856 & 29.1603 & 0.2927 & 1.00366 &    20.02\\
\end{tabular}\end{ruledtabular}
\caption{Thermodynamic functions for normal-liquid
  $^3$He at pressure $P=15\,\mbox{bar}$.}
\end{table*}

\begin{table*}[p]
\begin{ruledtabular}\begin{tabular}{ddddddd}
\multicolumn{1}{r}{$T$} &
\multicolumn{1}{r}{$S/R$} &
\multicolumn{1}{r}{$C_P/RT$} &
\multicolumn{1}{r}{$V$} &
\multicolumn{1}{r}{$-(\partial V/\partial P)_T$} &
\multicolumn{1}{r}{$\kappa$} &
\multicolumn{1}{r}{$\alpha$}\\
\multicolumn{1}{r}{$[\mbox{K}]$} &
\multicolumn{1}{r}{} &
\multicolumn{1}{r}{$[\mbox{K}^{-1}]$} &
\multicolumn{1}{r}{$[\mbox{cm}^3]$} &
\multicolumn{1}{r}{$[\mbox{cm}^3\mbox{bar}^{-1}]$} &
\multicolumn{1}{r}{$[10^{-2}\mbox{bar}^{-1}]$} &
\multicolumn{1}{r}{$[10^{-3}\mbox{K}^{-1}]$}\\
\hline
0     & 0      & 3.8176 & 27.7571 & 0.1971 & 0.70992 &    0\\
0.005 & 0.0191 & 3.8083 & 27.7571 & 0.1971 & 0.70992 &   -0.70\\
0.010 & 0.0381 & 3.7824 & 27.7569 & 0.1971 & 0.70992 &   -1.39\\
0.020 & 0.0755 & 3.6907 & 27.7563 & 0.1971 & 0.70993 &   -2.69\\
0.030 & 0.1117 & 3.5610 & 27.7554 & 0.1970 & 0.70994 &   -3.86\\
0.040 & 0.1466 & 3.4086 & 27.7542 & 0.1970 & 0.70995 &   -4.89\\
0.060 & 0.2115 & 3.0828 & 27.7510 & 0.1970 & 0.70999 &   -6.50\\
0.080 & 0.2701 & 2.7768 & 27.7471 & 0.1970 & 0.71005 &   -7.65\\
0.100 & 0.3228 & 2.5156 & 27.7426 & 0.1970 & 0.71013 &   -8.48\\
0.150 & 0.4341 & 1.9567 & 27.7302 & 0.1970 & 0.71033 &   -9.20\\
0.200 & 0.5211 & 1.5507 & 27.7174 & 0.1969 & 0.71051 &   -9.12\\
0.250 & 0.5912 & 1.2686 & 27.7051 & 0.1969 & 0.71075 &   -8.73\\
0.300 & 0.6493 & 1.0678 & 27.6933 & 0.1969 & 0.71107 &   -8.24\\
0.350 & 0.6989 & 0.9206 & 27.6822 & 0.1970 & 0.71149 &   -7.74\\
0.400 & 0.7420 & 0.8102 & 27.6719 & 0.1970 & 0.71201 &   -7.25\\
0.450 & 0.7803 & 0.7257 & 27.6622 & 0.1971 & 0.71264 &   -6.80\\
0.500 & 0.8149 & 0.6595 & 27.6530 & 0.1973 & 0.71337 &   -6.39\\
0.600 & 0.8757 & 0.5624 & 27.6365 & 0.1976 & 0.71508 &   -5.61\\
0.700 & 0.9283 & 0.4932 & 27.6221 & 0.1981 & 0.71704 &   -4.78\\
0.800 & 0.9748 & 0.4405 & 27.6102 & 0.1986 & 0.71917 &   -3.83\\
0.900 & 1.0168 & 0.3995 & 27.6011 & 0.1991 & 0.72144 &   -2.76\\
1.000 & 1.0550 & 0.3675 & 27.5950 & 0.1997 & 0.72384 &   -1.61\\
1.200 & 1.1238 & 0.3242 & 27.5929 & 0.2012 & 0.72912 &    0.85\\
1.400 & 1.1860 & 0.2996 & 27.6044 & 0.2029 & 0.73517 &    3.32\\
1.600 & 1.2444 & 0.2860 & 27.6294 & 0.2050 & 0.74214 &    5.72\\
1.800 & 1.3007 & 0.2781 & 27.6674 & 0.2075 & 0.75012 &    8.02\\
2.000 & 1.3558 & 0.2731 & 27.7180 & 0.2104 & 0.75920 &    10.24\\
2.500 & 1.4901 & 0.2647 & 27.8980 & 0.2196 & 0.78725 &    15.62\\
\end{tabular}\end{ruledtabular}
\caption{Thermodynamic functions for normal-liquid
  $^3$He at pressure $P=20\,\mbox{bar}$.}
\end{table*}

\begin{table*}[p]
\begin{ruledtabular}\begin{tabular}{ddddddd}
\multicolumn{1}{r}{$T$} &
\multicolumn{1}{r}{$S/R$} &
\multicolumn{1}{r}{$C_P/RT$} &
\multicolumn{1}{r}{$V$} &
\multicolumn{1}{r}{$-(\partial V/\partial P)_T$} &
\multicolumn{1}{r}{$\kappa$} &
\multicolumn{1}{r}{$\alpha$}\\
\multicolumn{1}{r}{$[\mbox{K}]$} &
\multicolumn{1}{r}{} &
\multicolumn{1}{r}{$[\mbox{K}^{-1}]$} &
\multicolumn{1}{r}{$[\mbox{cm}^3]$} &
\multicolumn{1}{r}{$[\mbox{cm}^3\mbox{bar}^{-1}]$} &
\multicolumn{1}{r}{$[10^{-2}\mbox{bar}^{-1}]$} &
\multicolumn{1}{r}{$[10^{-3}\mbox{K}^{-1}]$}\\
\hline
0     & 0      & 4.0499 & 26.8659 & 0.1620 & 0.60283 &    0\\
0.005 & 0.0202 & 4.0386 & 26.8659 & 0.1620 & 0.60283 &   -0.71\\
0.010 & 0.0404 & 4.0072 & 26.8657 & 0.1620 & 0.60284 &   -1.42\\
0.020 & 0.0799 & 3.8972 & 26.8652 & 0.1620 & 0.60285 &   -2.75\\
0.030 & 0.1182 & 3.7436 & 26.8643 & 0.1620 & 0.60287 &   -3.95\\
0.040 & 0.1547 & 3.5658 & 26.8630 & 0.1620 & 0.60289 &   -5.00\\
0.060 & 0.2223 & 3.1956 & 26.8599 & 0.1620 & 0.60296 &   -6.66\\
0.080 & 0.2828 & 2.8597 & 26.8560 & 0.1620 & 0.60306 &   -7.88\\
0.100 & 0.3370 & 2.5617 & 26.8515 & 0.1620 & 0.60320 &   -8.79\\
0.150 & 0.4493 & 1.9626 & 26.8391 & 0.1620 & 0.60354 &   -9.46\\
0.200 & 0.5363 & 1.5421 & 26.8264 & 0.1620 & 0.60387 &   -9.40\\
0.250 & 0.6058 & 1.2547 & 26.8140 & 0.1620 & 0.60425 &   -9.07\\
0.300 & 0.6632 & 1.0525 & 26.8021 & 0.1621 & 0.60468 &   -8.66\\
0.350 & 0.7119 & 0.9057 & 26.7908 & 0.1621 & 0.60518 &   -8.24\\
0.400 & 0.7544 & 0.7964 & 26.7800 & 0.1622 & 0.60576 &   -7.85\\
0.450 & 0.7920 & 0.7130 & 26.7698 & 0.1623 & 0.60641 &   -7.49\\
0.500 & 0.8260 & 0.6476 & 26.7600 & 0.1625 & 0.60713 &   -7.15\\
0.600 & 0.8856 & 0.5502 & 26.7417 & 0.1628 & 0.60875 &   -6.49\\
0.700 & 0.9369 & 0.4792 & 26.7254 & 0.1632 & 0.61052 &   -5.73\\
0.800 & 0.9819 & 0.4244 & 26.7112 & 0.1636 & 0.61236 &   -4.84\\
0.900 & 1.0221 & 0.3816 & 26.6996 & 0.1640 & 0.61426 &   -3.82\\
1.000 & 1.0586 & 0.3485 & 26.6909 & 0.1645 & 0.61622 &   -2.70\\
1.200 & 1.1235 & 0.3045 & 26.6827 & 0.1655 & 0.62033 &   -0.37\\
1.400 & 1.1817 & 0.2805 & 26.6869 & 0.1668 & 0.62484 &    1.95\\
1.600 & 1.2364 & 0.2676 & 26.7033 & 0.1682 & 0.62983 &    4.16\\
1.800 & 1.2891 & 0.2605 & 26.7312 & 0.1698 & 0.63535 &    6.27\\
2.000 & 1.3408 & 0.2560 & 26.7702 & 0.1717 & 0.64144 &    8.29\\
2.500 & 1.4667 & 0.2480 & 26.9138 & 0.1775 & 0.65936 &    13.08\\
\end{tabular}\end{ruledtabular}
\caption{Thermodynamic functions for normal-liquid
  $^3$He at pressure $P=25\,\mbox{bar}$.}
\end{table*}

\begin{table*}[p]
\begin{ruledtabular}\begin{tabular}{ddddddd}
\multicolumn{1}{r}{$T$} &
\multicolumn{1}{r}{$S/R$} &
\multicolumn{1}{r}{$C_P/RT$} &
\multicolumn{1}{r}{$V$} &
\multicolumn{1}{r}{$-(\partial V/\partial P)_T$} &
\multicolumn{1}{r}{$\kappa$} &
\multicolumn{1}{r}{$\alpha$}\\
\multicolumn{1}{r}{$[\mbox{K}]$} &
\multicolumn{1}{r}{} &
\multicolumn{1}{r}{$[\mbox{K}^{-1}]$} &
\multicolumn{1}{r}{$[\mbox{cm}^3]$} &
\multicolumn{1}{r}{$[\mbox{cm}^3\mbox{bar}^{-1}]$} &
\multicolumn{1}{r}{$[10^{-2}\mbox{bar}^{-1}]$} &
\multicolumn{1}{r}{$[10^{-3}\mbox{K}^{-1}]$}\\
\hline
0     & 0      & 4.1886 & 26.4036 & 0.1468 & 0.55603 &    0\\
0.005 & 0.0209 & 4.1762 & 26.4036 & 0.1468 & 0.55603 &   -0.73\\
0.010 & 0.0417 & 4.1415 & 26.4034 & 0.1468 & 0.55603 &   -1.44\\
0.020 & 0.0826 & 4.0207 & 26.4029 & 0.1468 & 0.55605 &   -2.80\\
0.030 & 0.1220 & 3.8529 & 26.4020 & 0.1468 & 0.55607 &   -4.03\\
0.040 & 0.1596 & 3.6603 & 26.4007 & 0.1468 & 0.55610 &   -5.10\\
0.060 & 0.2288 & 3.2643 & 26.3976 & 0.1468 & 0.55619 &   -6.81\\
0.080 & 0.2904 & 2.9113 & 26.3936 & 0.1468 & 0.55631 &   -8.07\\
0.100 & 0.3455 & 2.5870 & 26.3891 & 0.1468 & 0.55646 &   -9.03\\
0.150 & 0.4585 & 1.9660 & 26.3766 & 0.1469 & 0.55687 &   -9.68\\
0.200 & 0.5454 & 1.5377 & 26.3639 & 0.1469 & 0.55727 &   -9.63\\
0.250 & 0.6146 & 1.2476 & 26.3514 & 0.1470 & 0.55771 &   -9.33\\
0.300 & 0.6716 & 1.0446 & 26.3393 & 0.1470 & 0.55820 &   -8.96\\
0.350 & 0.7200 & 0.8981 & 26.3278 & 0.1471 & 0.55874 &   -8.58\\
0.400 & 0.7620 & 0.7895 & 26.3167 & 0.1472 & 0.55935 &   -8.22\\
0.450 & 0.7994 & 0.7068 & 26.3061 & 0.1473 & 0.56003 &   -7.91\\
0.500 & 0.8330 & 0.6416 & 26.2959 & 0.1475 & 0.56077 &   -7.61\\
0.600 & 0.8920 & 0.5437 & 26.2767 & 0.1478 & 0.56238 &   -7.00\\
0.700 & 0.9426 & 0.4714 & 26.2593 & 0.1481 & 0.56410 &   -6.27\\
0.800 & 0.9868 & 0.4153 & 26.2439 & 0.1485 & 0.56587 &   -5.39\\
0.900 & 1.0261 & 0.3714 & 26.2311 & 0.1489 & 0.56767 &   -4.37\\
1.000 & 1.0614 & 0.3377 & 26.2211 & 0.1493 & 0.56949 &   -3.27\\
1.200 & 1.1241 & 0.2934 & 26.2099 & 0.1502 & 0.57325 &   -0.98\\
1.400 & 1.1802 & 0.2698 & 26.2107 & 0.1513 & 0.57726 &    1.28\\
1.600 & 1.2328 & 0.2575 & 26.2231 & 0.1525 & 0.58160 &    3.43\\
1.800 & 1.2835 & 0.2508 & 26.2466 & 0.1539 & 0.58630 &    5.47\\
2.000 & 1.3333 & 0.2466 & 26.2804 & 0.1554 & 0.59136 &    7.42\\
2.500 & 1.4546 & 0.2391 & 26.4088 & 0.1600 & 0.60579 &   12.01\\
\end{tabular}\end{ruledtabular}
\caption{Thermodynamic functions for normal-liquid
  $^3$He at pressure $P=28\,\mbox{bar}$.\label{lasttable}}
\end{table*}

\begin{table*}[p]
\begin{ruledtabular}\begin{tabular}{rdddddd}
\multicolumn{1}{r}{$P$} &
\multicolumn{1}{r}{$V$} &
\multicolumn{1}{r}{$\gamma/R$} &
\multicolumn{1}{r}{$\displaystyle\frac{\displaystyle\partial\gamma}{\displaystyle\partial P}/R$} &
\multicolumn{1}{r}{$\kappa(0,P)$}
\\
\multicolumn{1}{r}{$[\mbox{bar}]$} &
\multicolumn{1}{r}{$[\mbox{cm$^3$}]$} &
\multicolumn{1}{r}{$ $} &
\multicolumn{1}{r}{$[10^{-2}\mbox{bar}^{-1}]$} &
\multicolumn{1}{r}{$[10^{-2}\mbox{bar}^{-1}]$}
\\
\hline
 0 & 36.846 & 2.7411 & 7.1003 & 3.2737 \\
 1 & 35.739 & 2.8101 & 6.7192 & 2.8460 \\
 2 & 34.798 & 2.8758 & 6.4265 & 2.5032 \\
 3 & 33.986 & 2.9388 & 6.1884 & 2.2231 \\
 4 & 33.279 & 2.9996 & 5.9843 & 1.9905 \\
 5 & 32.656 & 3.0586 & 5.8046 & 1.7954 \\
 6 & 32.102 & 3.1158 & 5.6447 & 1.6306 \\
 7 & 31.605 & 3.1715 & 5.5025 & 1.4908 \\
 8 & 31.157 & 3.2259 & 5.3764 & 1.3715 \\
 9 & 30.749 & 3.2791 & 5.2650 & 1.2691 \\
10 & 30.374 & 3.3313 & 5.1670 & 1.1808 \\
11 & 30.030 & 3.3825 & 5.0810 & 1.1041 \\
12 & 29.710 & 3.4329 & 5.0059 & 1.0371 \\
13 & 29.412 & 3.4826 & 4.9404 & 0.9782 \\
14 & 29.134 & 3.5317 & 4.8835 & 0.9261 \\
15 & 28.872 & 3.5803 & 4.8343 & 0.8798 \\
16 & 28.625 & 3.6285 & 4.7920 & 0.8385 \\
17 & 28.392 & 3.6762 & 4.7558 & 0.8013 \\
18 & 28.170 & 3.7236 & 4.7252 & 0.7678 \\
19 & 27.959 & 3.7707 & 4.6995 & 0.7375 \\
20 & 27.757 & 3.8176 & 4.6782 & 0.7099 \\
21 & 27.564 & 3.8643 & 4.6610 & 0.6847 \\
22 & 27.379 & 3.9108 & 4.6476 & 0.6617 \\
23 & 27.202 & 3.9573 & 4.6375 & 0.6405 \\
24 & 27.031 & 4.0036 & 4.6304 & 0.6209 \\
25 & 26.866 & 4.0499 & 4.6263 & 0.6028 \\
26 & 26.707 & 4.0961 & 4.6247 & 0.5861 \\
27 & 26.553 & 4.1424 & 4.6256 & 0.5705 \\
28 & 26.404 & 4.1886 & 4.6287 & 0.5560 \\
29 & 26.259 & 4.2350 & 4.6340 & 0.5425 \\
\end{tabular}\end{ruledtabular}
\caption{Pressure dependence of the molar volume $V$, 
  linear coefficient of the specific heat $\gamma(P)$, 
  its derivative $\gamma'(P)$, and compressibility
  $\kappa$, for $T\to0$.   See also the discussion of the accuracy of
  $\left(\frac{\partial{V}}{\partial{T}}\right)_{\!{P}}$ 
  at low pressures in Sec.~IV.C and
  Ref.~\protect\onlinecite{footnote3}.\label{zerotemptable}}
\end{table*}


\begin{thebibliography}{1}
\bibitem{footnote1}
    In principle, one can use any sufficiently accurate set of data
    for the molar volume $V(T_0,P)$ at some fixed temperature $T_0$
    for this purpose.  However, since Greywall estimated his
    interpolation formula for $V(0.1\mbox{K},P)$ to be accurate within
    0.02 cm$^3$ (rms deviation) and used it to determine the volume of
    his experimental cell, we decided to employ his values of the
    molar volume, rather than other group's measurements of this
    quantity, thereby ensuring internal consistency.
\bibitem{footnote2}
        Greywall's interpolation formula $v_0(P)$ [Eq.~(\ref{v0def})] is a
    fifth-order polynomial in $P$ which parametrizes the data for the
    molar volume with sufficient accuracy
    (Ref.~\onlinecite{footnote1}). However, the calculation of the
    quantities $C_P$ $-$ $C_V$,
    $\left(\frac{\partial{C_P}}{\partial{P}}\right)_{\!{T}}$,
    $\kappa(T,P)$, and $\alpha(T,P)$ from $F(T,V)$ [Eq.~(\ref{ftv})]
    involves also the derivative of $v_0(P)$, $dv_0/dP$. This is then
    only a fourth-order polynomial in $P$ which is no longer
    sufficiently accurate.  Therefore we reparametrized $v_0(P)$, or
    rather the inverse function $p_0(V)$ [Eq.~(\ref{p0def})], by a
    polynomial of higher order.
\bibitem{footnote3}    
    We note that the derivatives
    $\left(\frac{\partial{V}}{\partial{P}}\right)_{\!{T_0}}$ and
    $\left(\frac{\partial{P}}{\partial{V}}\right)_{\!{T_0}}$, obtained
    from Greywall's interpolation formula for $V(T_0,P)$
    [Eq.~(\ref{v0def})], cannot be expected to be as accurate as
    $V(T_0,P)$ itself. Since the variances of the coefficients in
    Eq.~(\ref{v0def}) are not known it is not possible to determine
    the accuracy of these derivatives with precision. However, the
    largest part of the problem was already removed by our
    reparametrization of $p_0(V)$ [Eq.~(\ref{p0def})]
    (Ref.~\onlinecite{footnote2}), which eliminates the shortcomings
    of fitting $V(T_0,P)$ by a rather low-order polynomial in $P$. The
    remaining systematic error in the derivative of the interpolation
    formula (\ref{v0def}) will be largest at the two ends of the
    fit curve $v_0(P)$, i.e., at $P$ $\lesssim$ 2~bar and $P$
    $\gtrsim$ 27~bar. Since $v_0(P)$ is most rapidly changing at small
    $P$ we deduce that the error in
    $\left(\frac{\partial{V}}{\partial{P}}\right)_{\!{T}}$ is largest
    at $P$ $\lesssim$ 2~bar.
\bibitem{footnote4}    
    In view of the unknown systematic error
    (Refs.~\onlinecite{footnote2,footnote3}) of
    $\left(\frac{\partial{V}}{\partial{P}}\right)_{\!{T}}$ and its
    implications for the compressibility especially at low pressures
    (see Sec.~IV.C), 
    our values for $C_P(T,P$ $=$ $0)$, obtained from $C_P$ $-$ $C_V$
    $=$
    $VT\kappa\left(\frac{\partial{P}}{\partial{T}}\right)_{\!{V}}^2$
    [Eq.~(\ref{cpcv})], may be subject to an error of at most a few
    percent at high temperatures.
\bibitem{footnote5}
    Since $\alpha$ is proportional to $\kappa$ [Eq.~(\ref{alphadef})]
    any inaccuracy in $\kappa$ (see discussion above) will enter also
    in $\alpha$.
\end{thebibliography}
\end{document}